\documentclass[12pt]{article}
\usepackage{amssymb}
\usepackage{amsfonts}
\usepackage{youngtab}
\usepackage{amsmath,amssymb,latexsym,cite}
\usepackage{amsthm}
\usepackage{axodraw}
\usepackage{cite}
\usepackage[]{hyperref}
\input xy
\xyoption{all}

\def\sqr#1#2{{\vcenter{\vbox{\hrule height.#2pt
            \hbox{\vrule width.#2pt height#1pt \kern#1pt
                  \vrule width.#2pt}\hrule height.#2pt}}}}

\def\square
 {\mathop{\mathchoice{\sqr{12}{15}}{\sqr{9}{12}}
{\sqr{6.3}{9}}{\sqr{4.5}{9}}}}

\def\sqra#1#2#3{{\vcenter{\vbox{\hrule height.#2pt
            \hbox{\vrule width.#2pt height#1pt \kern5pt 
#3
                  \vrule width.#2pt}\hrule height.#2pt}}}}

\def\tsquare#1{\sqra{12}{15}{#1}}

\numberwithin{equation}{section}

\begin{document}

\vspace*{0.5in}

\begin{center}

{\large\bf Undoing decomposition}

\vspace{0.2in}

Eric Sharpe

Dep't of Physics\\
Virginia Tech\\
850 West Campus Dr.\\
Blacksburg, VA  24061\\

{\tt ersharpe@vt.edu}

$\,$

\end{center}

In this paper we discuss gauging one-form symmetries in
two-dimensional theories.  The existence of a global one-form symmetry
in two dimensions typically signals a violation of cluster decomposition --
an issue resolved by the observation that such theories
decompose into disjoint unions, a result that has been applied to,
for example,
Gromov-Witten theory and gauged linear sigma model phases.  
In this paper we describe
how gauging one-form symmetries in two-dimensional theories
can be used to select particular
elements of that disjoint union, effectively undoing decomposition.
We examine such gaugings explicitly in examples involving orbifolds,
nonsupersymmetric pure Yang-Mills theories, and supersymmetric gauge
theories in two dimensions.
Along the way, we learn explicit
concrete details of the topological configurations that
path integrals sum over when gauging a one-form symmetry,
and we also uncover `hidden' one-form symmetries.

\begin{flushleft}
November 2019
\end{flushleft}

\newpage

\tableofcontents

\newpage

\section{Introduction}

This paper is devoted to gauging one-form symmetries in two-dimensional
theories.  Two-dimensional theories with global one-form symmetries have been
studied for a number of years, see for example
\cite{Pantev:2005zs,Pantev:2005rh,Pantev:2005wj,Hellerman:2006zs,Caldararu:2007tc},
which discussed a variety of examples in orbifolds and gauge theories,
including, for example, analogues of the
supersymmetric ${\mathbb P}^n$ model, and how these theories are different
from ordinary theories via theta angle periodicities, massless spectra,
partition functions, anomalies,
quantum cohomologies, and mirrors.  (Lattice gauge theories with
analogous properties had been studied even earlier.)  
In particular,
any two-dimensional orbifold or
gauge theory in which a finite subgroup of the gauge
group acts trivially can exhibit a one-form symmetry, under which
one 
modifies any gauge bundle by tensoring in a bundle whose structure
group is in the trivially-acting subgroup to get a different nonperturbative
sector that is symmetric with respect to the original one.
(These
theories can also be understood as sigma models on generalized spaces known
as gerbes, which geometrically admit one-form symmetries,
though we shall
not emphasize that perspective in this paper.)  

One of the properties of these theories is that they violate
cluster decomposition (as can be seen, for example, from the multiplicity
of dimension zero operators), but they do so in the mildest possible way.
Specifically, such theories `decompose' into disjoint unions of theories,
a result described in \cite{Hellerman:2006zs} as the `decomposition conjecture.'

The decomposition conjecture has been checked in a wide variety of 
ways and in numerous two-dimensional examples.  We list here a few highlights:
\begin{itemize}
\item In orbifolds, decomposition reproduces multiloop partition functions,
correlation functions, and massless spectra \cite{Pantev:2005rh,Hellerman:2006zs}.
\item In gauged linear sigma models (GLSMs), decomposition reproduces
quantum cohomology rings and is manifestly visible in mirrors
\cite{Pantev:2005zs,Hellerman:2006zs,Gu:2018fpm,Chen:2018wep,Gu:2019zkw}.
\item In two-dimensional
nonsupersymmetric pure Yang-Mills theories, decomposition
reproduces partition functions and correlation functions of Wilson loops
\cite{Sharpe:2014tca}.
\item In supersymmetric two-dimensional gauge theories,
decomposition reproduces partition functions via splitting 
lattices \cite{Sharpe:2014tca}.
\item In K theory, decomposition reproduces the structure of the K theory
groups, and visibly illustrates how D-brane charges split into charges for
two distinct summands \cite{Hellerman:2006zs}.  Derived categories decompose
similarly \cite{Hellerman:2006zs}.
Ext groups also are only nonzero between complexes corresponding to
the same component, corresponding to the fact that open strings endpoints must 
lie on the same connected component in a disjoint union.
\end{itemize}
Other examples of decomposition
and computations checking decomposition in two-dimensional theories
are outlined in
\cite{Hellerman:2006zs,Sharpe:2014tca}.

The decomposition conjecture \cite{Hellerman:2006zs} makes a prediction
for Gromov-Witten theory, namely that Gromov-Witten invariants of gerbes
should be equivalent to Gromov-Witten invariants of disjoint unions of
ordinary spaces.  This was checked and proven rigorously in the
mathematics literature, see e.g.
\cite{ajt1,ajt2,ajt3,tseng1,gt1,xt1}, reproducing expectations from physics.

Another application of decomposition was to understand phases of
certain gauged linear sigma models \cite{Caldararu:2007tc}.
Briefly, in certain theories, locally in a Born-Oppenheimer approximation one
has a ${\mathbb Z}_2^{(1)}$ one-form symmetry, so that the theory decomposes.
This one-form symmetry is broken along a codimension-one locus, about which
there are nontrivial Berry phases.  This results in a geometric interpretation
as a branched double cover.  This trick has been utilized since in
for example 
\cite{Sharpe:2012ji,Addington:2012zv,Hori:2011pd,Halverson:2013eua,Hori:2013gga},
and see also \cite{Wong:2017cqs} for
a recent summary.

A discussion of decomposition in two-dimensional theories as a limit
of dualities in three-dimensional theories is given in 
\cite{Aharony:2017adm}.

In this paper, we will see explicitly 
how gauging such discrete one-form symmetries
in two-dimensional theories can `undo' decomposition, by projecting onto
components of the decomposition.  Some highlights include:
\begin{itemize}
\item We will make a prediction for the topological classes that path
integrals should sum over when gauging one-form symmetries.  Specifically,
it appears that path integrals only sum over `banded' gerbes, not more
general gerbes, when gauging one-form symmetries.  We will also see
explicitly how gauge theories in a sector
of a nontrivial gerbe are modified.
\item We will uncover `hidden' one-form symmetries.  Specifically, we will
see in examples that in addition to the `obvious' one-form symmetries
that arise in gauge theories with matter that is invariant under
a subgroup of the gauge group, there can be additional one-form symmetries
that do not correspond to trivially-acting gauge subgroups.
\end{itemize}

We should mention that there has recently been a great deal of
interest in gauging one-form symmetries in other dimensions,
see for example 
\cite{Bolognesi:2019fej,Anber:2019nze,Tanizaki:2017bam,Tanizaki:2018wtg}.

We begin in section~\ref{sect:review} by reviewing work on decomposition
of two-dimensional theories with one-form symmetries.
In section~\ref{sect:oneform-disjoint} we review how sigma models on
disjoint unions of spaces naturally admit one-form symmetries, to help
illustrate the connection between decomposition and one-form symmetries
in two-dimensional theories.  In section~\ref{sect:existence-orbifolds}
we turn to one-form symmetries in two-dimensional orbifolds, and discuss
how discrete torsion and various modified group actions can obstruct
the existence of a one-form symmetry that would otherwise arise from
a trivial group action.  In section~\ref{sect:gauging} we explicitly
describe the gauging of one-form symmetries in orbifolds, and discuss
how such one-form gauging can `undo decomposition' by projecting onto
particular factors in the decomposition.
In section~\ref{sect:exs-orbs} we discuss a number of explicit orbifold
examples, to make clear both the decomposition of a two-dimensional
theory with a one-form symmetry, as well as the detailed structure of
gauging that one-form symmetry and how the decomposition summands are
recovered via gauging.
We conclude that discussion of orbifold examples by outlining two-dimensional
Dijkgraaf-Witten theory in this language, which is just a theory of
orbifolds of a point.

In section~\ref{sect:pure-ym} we turn our attention to gauge theories
with non-finite gauge groups, beginning with the case of pure Yang-Mills
theory in two dimensions, for which there exists exact expressions
for partition functions and correlation functions.
We review how decomposition arises in two-dimensional pure Yang-Mills
theories with center symmetry, and how the corresponding one-form
symmetry can be explicitly gauged, recovering partition functions of
the decomposition summands.  In section~\ref{sect:susygauge} we 
discuss two-dimensional (2,2) supersymmetric gauge theories.
In particular, we discuss the effect of gauging one-form symmetries
in a family of gerby generalizations of the supersymmetric
${\mathbb P}^n$ model, and how various physical features of those
theories and their one-form gauging can be seen in their mirrors,
mirrors to both abelian and nonabelian gauge theories.
We also discuss partition functions for such supersymmetric gauge
theories obtained via supersymmetric localization, and explicitly illustrate
the one-form-symmetry gauging at the level of such partition functions,
verifying that the effect is to select summands in decomposition.
In section~\ref{sect:Ktheory} we discuss how these matters can be
seen in open string charges as computed in K theory.

Finally, in appendix~\ref{app:stacks} we discuss how these phenomena
can be understood geometrically in terms of generalized spaces known
as stacks.  In particular, although we will not emphasize this point
of view, the theories described in this paper also form examples of
sigma models on special stacks known as gerbes, which
geometrically admit one-form symmetries.  This is one geometric
way of understanding the presence of global one-form symmetries in
these two-dimensional theories:  just as a sigma model on a space with
an action of an ordinary group $G$ itself has a global symmetry group $G$,
so too do sigma models on generalized spaces with actions of $BG = G^{(1)}$
admit global $BG = G^{(1)}$ symmetries.  It is also worth noting that 
the results in this paper implicitly are making
predictions for the Gromov-Witten invariants of higher stacks.

\section{Review of decomposition}
\label{sect:review}

In orbifolds and two-dimensional gauge theories with a discrete one-form
symmetry,
cluster decomposition is typically violated.  This issue was studied
in \cite{Hellerman:2006zs}, where it was argued that the theories
are equivalent to disjoint unions, which also violate cluster decomposition,
but do so in a controllable fashion.

The paper \cite{Hellerman:2006zs} focused on orbifolds and gauged sigma models
in which a subgroup of the orbifold or gauge group acts trivially
(hence, the theory admits a discrete one-form symmetry).  
Specifically, suppose one has a $G$ orbifold 
or $G$-gauged sigma model of a space $X$ in which
a finite normal subgroup $K \subseteq G$ acts trivially.
The theory then has\footnote{
One-form symmetries are only defined for abelian groups.
If $K$ is abelian, then the theory has a $K^{(1)}$ one-form symmetry.
If $K$ is not abelian, then $Z(K)^{(1)}$ acts on the fibers of the $K$-gerbe.
} a $B Z(K) = Z(K)^{(1)}$ one-form symmetry,
where $Z(K)$ denotes the center of $K$.
(The one-form symmetry acts by tensoring any gauge bundle with a $B Z(K)$
bundle -- if all the matter is invariant under $Z(K)$, then, this
symmetrically permutes nonperturbative sectors, defining a one-form
symmetry, barring obstructions defined by giving non-symmetric
phases to different sectors related in this fashion, as in
section~\ref{sect:existence-orbifolds}.)

Let $\hat{K}$ denote the set of irreducible representations of $K$.
$\hat{K}$ admits a natural action of the group $H \equiv G/K$.
Then, the central claim of \cite{Hellerman:2006zs} is that the
$G$ orbifold or gauged sigma model is equivalent to
a sigma model on $Y \equiv [(X \times \hat{K})/H]$.

The space $Y$ will have multiple disjoint components, as many as
orbits of $H$ on $\hat{K}$.  If $H_1, \cdots, H_n$ are the stabilizers
in $H$ of the various orbits, then
\begin{equation}
Y \: = \: \coprod_i [X/H_i],
\end{equation}
and each component has a natural $B$ field, as described in
\cite{Hellerman:2006zs}.    
See 
e.g. \cite{Sharpe:2014tca} for analogous statements for other two-dimensional
gauge theories, not necessarily with a geometric interpretation.
Such theories also admit a decomposition into disjoint QFTs.
The result above has a number of applications, e.g. to Gromov-Witten theory
and to understanding phases of gauged linear sigma models, as was
outlined in the introduction.

Now, this picture can be simplified.  Quotients in which a nontrivial
subgroup acts trivially define generalized spaces known as gerbes,
Gerbes are better known in connection with $B$ fields, but can also
be interpreted as analogues of spaces, special cases of stacks,
admitting metrics, spinors, bundles, gauge fields, and so forth, just
like an ordinary manifold, and more to the point, admit one-form
symmetries, just as an ordinary space might admit an ordinary group
of symmetries.  In any event, there are special classes of gerbes
known as `banded gerbes,' for which decomposition simplifies.
Briefly, a $G$ gerbe over a space $M$
is said to be banded if it is classified by $H^2(M, C^{\infty}(G))$.
(More general gerbes have a more complicated classification.)
For examples, if $G=U(1)$, it is the banded gerbes whose connections are
$B$ fields.  (See \cite[section 3]{Hellerman:2006zs} 
for more information on the distinction 
between banded and non-banded gerbes.)

In the special case that the quotient $[X/G]$ defines a banded $K$-gerbe,
the description of decomposition simplifies.  
In this case, the $H$ action on $\hat{K}$ is
trivial, hence $[ ( X \times \hat{K}) / H] \cong [X/H] \times \hat{K}$,
and so there are as many components as elements of $\hat{K}$.  Furthermore,
the flat $B$ field on each component is determined by the image of
the characteristic class of the gerbe under a map defined by 
the corresponding irreducible representation $\rho \in \hat{K}$:
\begin{equation}
H^2([X/H], Z(K) ) \: \stackrel{\rho}{\longrightarrow} \: 
H^2([X/H], U(1)).
\end{equation}

Briefly, in this paper we will argue that decomposition can be undone
by gauging the corresponding one-form symmetries.  There can be multiple
such one-form symmetry actions on a given theory:  we can weight different
sectors in the path integral by phases, and use those phases to
select amongst the different components.  As we shall see in more detail
in examples later, when we gauge a one-form symmetry, the path integral
sums over banded gerbes on the worldsheet.  
We can weight the different gerbe sectors
by an analogue of a discrete theta angle, a phase factor in the path
integral of the form
$\exp\left( w_2(\xi) \right)$,
where $\xi$ is the gerbe on the worldsheet appearing in the sum,
and $\exp(w_2(\xi))$ is a phase determined by the characteristic
class of the gerbe on the worldsheet.
In terms of decomposition, for any irreducible representation
$\rho$ of $G$,
we can assign a phase to a banded $G$ gerbe on the worldsheet $\Sigma$
as the image of
\begin{equation}
H^2(\Sigma, G) \: \stackrel{\rho}{\longrightarrow} \:
H^2(\Sigma, U(1)).
\end{equation}
We shall see that for this phase factor, we recover the component
corresponding to $\rho$.
(In this paper, we focus on gauging one-form symmetries in banded gerbes,
and leave more general discussions for later work.)

\section{One-form symmetries in disjoint unions}
\label{sect:oneform-disjoint}

In this section we will discuss how sigma models on disjoint unions
admit (global) discrete one-form symmetries, 
acting on discrete Fourier transforms
of projection operators and domain walls.

Suppose $X$ is a disjoint union of $k$ Calabi-Yau's, denoted $X_i$:
\begin{equation}
X \: = \: \coprod_{i=0}^{k-1} X_i.
\end{equation}
A sigma model on $X$ will admit a collection of projection operators,
projecting onto states associated with each summand $X_i$.
These are constructed as linear combinations of the dimension zero operators
in the theory (of which there will be one for every connected component
$X_i$.)  (See e.g. \cite{Hellerman:2006zs} for further details.)

Let $\Pi_i$ denote a projection operator in the CFT of a sigma
model on $X$, projecting onto states corresponding to $X_i$,
where $i \in \{0, \cdots, k-1 \}$.
Discrete Fourier transforms of these projection operators 
form the elements of the group ${\mathbb Z}_k$.  For example,
\begin{equation}
1 \: = \: \sum_{j=0}^{k-1} \Pi_j,
\end{equation}
as projecting onto all possibilites is the same as the identity operator.
A general discrete Fourier transform takes the form
\begin{equation}
A(p) \: = \: \sum_{j=0}^{k-1} \Pi_j \, \xi^{jp},
\end{equation}
where
\begin{equation}
\xi \: \equiv \: \exp\left( \frac{2 \pi i}{k} \right).
\end{equation}
Note that if we define
\begin{equation}
z \: \equiv \: A(1) \: = \:
\sum_{j=0}^{k-1} \Pi_j \, \xi^{j},
\end{equation}
then it is straightforward to show that
\begin{equation}
z^p \: = \: A(p),
\end{equation}
using the property
\begin{equation}
\Pi_m \Pi_n \: = \: \left\{ \begin{array}{cl}
0 & m \neq n, \\
\Pi_m & m = n
\end{array} \right.
\end{equation}
of projection operators.  Furthermore, these linear combinations have the
multiplications of the group ${\mathbb Z}_k$.  For example, it is
straightforward to compute that
\begin{eqnarray}
z^p z^q & = & \left( \sum_{j=0}^{k-1} \Pi_j \, \xi^{p j}
\right)
\left( \sum_{\ell=0}^{k-1} \Pi_{\ell} \, \xi^{q \ell}
 \right),
\\
& = & \sum_{j=0}^{k-1} \Pi_j \, \xi^{(p+q) j},
\\
& = & z^{p+q},
\end{eqnarray}
and also that $z^0 = 1$.
As a result, these linear combinations of
projection operators form the elements of the group
${\mathbb Z}_k$.  Under an $SL(2,{\mathbb Z})$ transformation,
these projection operators become, in effect, domain walls on the
worldsheet.

Now, consider the case that the worldsheet of the sigma model is
$T^2$, for simplicity.  One can have domain walls along both the
spacelike and timelike directions, giving rise to boundary conditions
in close analogy with orbifolds.  We have just seen that linear combinations
of projection operators form elements of the group ${\mathbb Z}_k$,
so we can think of a pair of domain walls on the worldsheet as linear
combinations of sectors closely analogous to orbifold twisted sectors,
of the form
\begin{equation}
{\scriptstyle z^m} \square_{z^n}.
\end{equation}
An element
\begin{equation}
\left[ \begin{array}{cc} a & b \\ c & d \end{array} \right]
\: \in \: SL(2,{\mathbb Z})
\end{equation}
acts on such a sector by mapping
\begin{equation}
( z^m, z^n ) \: \mapsto \: ( z^{am + bn}, z^{cm + dn} ).
\end{equation}
More pertinently, to each such twisted sector, we can associate
a ${\mathbb Z}_k$ bundle on the worldsheet.

Now, we are ready to discuss the ${\mathbb Z}_k^{(1)}$ one-form
symmetry.  Given a ${\mathbb Z}_k$ bundle, we simply tensor that
bundle with the bundle corresponding to a linear combination of
projection operators/domain walls, to get another such.
For example, consider the case of ${\mathbb Z}_2$,
and take the worldsheet to be $T^2$.
The ${\mathbb Z}_2$ bundle corresponding to
\begin{equation}
{\scriptstyle z} \square_z
\end{equation}
maps, for example,
\begin{eqnarray}
{\scriptstyle 1} \square_1 & \mapsto &
{\scriptstyle z} \square_z,
\\
{\scriptstyle 1} \square_z & \mapsto &
{\scriptstyle z} \square_1,
\\
{\scriptstyle z} \square_z & \mapsto &
{\scriptstyle 1} \square_1.
\end{eqnarray}
Thus, we see in this case that there is a well-defined
${\mathbb Z}_2^{(1)}$ one-form action, which permuts the various
projection operators and domain walls of the sigma model on
the disjoint union.

In passing, note we have not assumed any symmetry between the various
connected components $X_i$ of $X$, only that each is nonempty.  
This formal argument applies to
any space with multiple disconnected (nonempty) components, regardless of the
components.  Later in this paper we will study two-dimensional 
theories obtained by
gauging groups with trivially-acting subgroups, which will be equivalent
to disjoint unions of theories but with components related by various
symmetries, a special case of the picture presented in this section.

So far, we have established that a sigma model on a space $X$ with
$k$ connected components has a ${\mathbb Z}_k^{(1)}$ one-form symmetry,
an action of ${\mathbb Z}_k^{(1)}$.
We should pause at this point to observe that this action is not unique.
Suppose for example $X$ decomposes into four components $X_i$,
so that, from the analysis above, a sigma model on $X$ admits
an action of ${\mathbb Z}_4^{(1)}$.
Let us illustrate two different sets of ways to get actions of different
one-form groups:
\begin{itemize}
\item First, group the four $X_i$ into two pairs $Y_0$, $Y_1$,
so that $X = \coprod Y_a$ and each $Y_a$ itself can be decomposed.
Applying the same argument as above, there is a 
${\mathbb Z}_2^{(1)}$ action, that acts by interchanging discrete Fourier
transforms of domain walls and projectors onto those two components
$Y_0$, $Y_1$.  By further decomposing the $Y_a$ and repeating,
one sees that ${\mathbb Z}_2^{(1)} \times {\mathbb Z}_2^{(1)}$ also
acts on a sigma model on $X$.
\item Second, write $Y = X_2 \coprod X_3$ and consider the decomposition
\begin{equation}
X \: = \: X_0 \coprod X_1 \coprod Y.
\end{equation}
From the analysis above, we see that there is an action of
${\mathbb Z}_3^{(1)}$ that interchanges discrete Fourier transforms of
domain walls and projectors onto $X_0$, $X_1$, and $Y$.
\end{itemize}
Thus, for a sigma model on $X$ with four disconnected components,
we have derived actions of ${\mathbb Z}_4^{(1)}$,
${\mathbb Z}_2^{(1)}$, ${\mathbb Z}_2^{(1)} \times
{\mathbb Z}_2^{(1)}$, and ${\mathbb Z}_3^{(1)}$.
Thus, for theories of this form, one-form symmetry groups are not unique,
and which to use will vary depending upon the application.

In passing, note that this non-uniqueness is not specific to one-form
symmetries, and also arises in theories with ordinary group symmetries.
Consider for example an $SO(3)$ WZW model in two dimensions.
It admits a (symmetric) action of $SO(3)$, but also admits a (symmetric)
action of $SU(2)$, as the ${\mathbb Z}_2$ center simply factors out.
For that matter, there is also a (symmetric) action of $U(1)$,
as $U(1)$ is a maximal torus of $SO(3)$.

Next, we shall consider gauging one-form symmetries.  We will just
outline basics in this section, and will consider this in more detail
in later sections.  Briefly, we claim that by gauging the one-form
symmetries described above, one can pick out particular summands,
particular connected components, in a decomposition.
(Our discussion will anticipate and cite methodology that will
be justified later in section~\ref{sect:gauging}.  As a result,
readers may wish to skip the remainder of this section for the moment
and only return after reading section~\ref{sect:gauging}.)

To this end, we return to the example of a sigma model on a disjoint
union of $k$ spaces $X_n$, which has a $B {\mathbb Z}_k = {\mathbb Z}_k^{(1)}$
one-form symmetry.
It is straightforward to show that
\begin{equation}
\Pi_n \: = \: \frac{1}{k} \sum_{j=0}^{k-1} z^j \xi^{-n j},
\end{equation}
and 
any given summand in the disjoint union has partition function
\begin{equation}  \label{eq:proj-component}
{\rm CFT}(X_n) \: = \:
{\scriptstyle \Pi_n} \square_{\Pi_n} \: = \: 
\frac{1}{k^2} \sum_{p=0}^{k-1} \sum_{q=0}^{k-1} \xi^{-pn} \xi^{-qn}
{\scriptstyle z^p} \square_{z^q}.
\end{equation}
We claim that this partition function, for a given component, can be
obtained by gauging the ${\mathbb Z}_k^{(1)}$ one-form symmetry.

When we gauge a $G^{(1)}$ one-form symmetry, in the partition function,
\begin{enumerate}
\item we first sum over banded\footnote{
We will justify summing only over banded gerbes, and not more general
gerbes, later in this paper.
} $G$-gerbes,
\item then in each $G$-gerbe sector, we sum over field configurations twisted
by the gerbe, multiplied by a gerbe-dependent phase.
\end{enumerate}
Here, this means that the partition function schematically has the form
\begin{equation}
Z \: = \: \frac{1}{k} \sum_{z \in {\mathbb Z}_k}
\epsilon(z) \cdots,
\end{equation}
where $z$ defines the characteristic class of the banded gerbe,
$\epsilon(z)$ is the gerbe-dependent phase, and the $\cdots$ indicates
the path integral over gerbe-twisted field configurations.
When constructing the partition function of an orbifold on worldsheet $T^2$,
as we shall describe in detail in section~\ref{sect:gauging},
in a $z$-gerbe twisted sector, instead
of summing over commuting pairs of group elements, one sums over
pairs $(g,h)$ obeying $gh = hgz$.  Here, since ${\mathbb Z}_k$ is abelian,
all group elements commute, so $gh = hgz$ only has solutions when $z=1$,
so the $z$-gerbe twisted sectors are empty except in the case $z=1$.
Furthermore, for $z=1$, $\epsilon(z) = 1$, as we argue later.
As a result, all that is left, aside from the factor of $1/k$, is
an ordinary ${\mathbb Z}_k$ orbifold partition function for the $\cdots$
in the schematic description above.  That ${\mathbb Z}_k$ orbifold 
is equivalent to the quantum symmetry orbifold described in
\cite[section 8.5]{Ginsparg:1988ui}.  Adding the extra factor of $1/k$
(from the remnant of the sum over ${\mathbb Z}_k$ gerbes),
we get that the partition function of a sigma model on the disjoint
union after gauging the ${\mathbb Z}_k^{(1)}$ one-form symmetry is
\begin{equation}
\frac{1}{k^2} \sum_{p=0}^{k-1} \sum_{q=0}^{k-1} \xi^{-pn} \xi^{-qn}
{\scriptstyle z^p} \square_{z^q},
\end{equation}
which precisely matches the partition function of a component $X_n$
as above in equation~(\ref{eq:proj-component}).

\section{Existence of one-form symmetries in orbifolds}
\label{sect:existence-orbifolds}

We have just discussed one set of theories with global
one-form symmetries, namely, sigma models on
disjoint unions.
As previously mentioned, another common source of theories with one-form
symmetries is a gauge theory (or orbifold) in which a subgroup of
the gauge group acts trivially on the matter fields.  In such a theory,
there is a one-form symmetry, which acts by permuting the nonperturbative
sectors (gauge bundles) by tensoring in bundles whose structure group
lies in the trivially-acting subgroup.  (These theories are related to
disjoint unions via decomposition, as reviewed in section~\ref{sect:review}.)

However, the one-form symmetry in such a gauge theory can be broken,
by giving different phases to nonperturbative sectors permuted
by the one-form symmetry,
and when the one-form symmetry is broken, so too is the 
decomposition prediction of \cite{Hellerman:2006zs}. 
In this section we will discuss specific examples in which
the one-form symmetries arising in gauge theories with trivially-acting
subgroups are broken.

\subsection{Discrete torsion}
\label{sect:dt}

One way to break one-form symmetries is to add phases which are
asymmetric between sectors related by tensoring by bundles associated
with the one-form symmetry.  Such phases are constrained by 
modular invariance, and a classic example in orbifolds
is that of discrete torsion \cite{Vafa:1986wx,Sharpe:2000ki}.
In this subsection,
we will look at some examples of how turning on discrete torsion
breaks one-form symmetries (and also decomposition), taken from
\cite[section 10]{Hellerman:2006zs}.

First, consider the orbifold $[X/ {\mathbb Z}_2 \times
{\mathbb Z}_2]$, where the first ${\mathbb Z}_2$ acts
trivially but the second ${\mathbb Z}_2$
acts nontrivially on $X$.  This was studied in
\cite[section 10.1]{Hellerman:2006zs}.
Since 
\begin{equation}
H^2( {\mathbb Z}_2 \times {\mathbb Z}_2, U(1) ) \: = \: 
{\mathbb Z}_2,
\end{equation}
this orbifold does admit the possibility of turning on
(exactly one choice of) discrete torsion.

If we do not turn on discrete torsion, then the one-loop twisted
sectors are invariant under tensoring with arbitrary
${\mathbb Z}_2$ bundles, which simply permuts the various
twisted sectors.  As a result, the theory admits a
${\mathbb Z}_2^{(1)}$
one-form symmetry, and as argued in \cite{Hellerman:2006zs},
the theory decomposes into a sigma model
on a disjoint union of two copies of $[X/{\mathbb Z}_2]$.

Now, let us modify this orbifold by turning on 
discrete torsion, which will weight the various one-loop twisted
sectors differently and thereby break the one-form symmetry above.  

Let $a$ denote the generator of the first (trivial) ${\mathbb Z}_2$,
and $b$ the generator of the second.  
Discrete torsion acts by multiplying the $T^2$ twisted sectors
\begin{equation}
{\scriptstyle a} \square_b, \: \: \:
{\scriptstyle a} \square_{ab}, \: \: \:
{\scriptstyle b} \square_{ab}
\end{equation}
by $-1$.

First, the 
\begin{equation}
{\scriptstyle 1} \square_1
\end{equation}
one-loop sector in the effective $[X/{\mathbb Z}_2]$ orbifold emerges
from any of the sectors
\begin{equation}
{\scriptstyle 1, a} \square_{1, a},
\end{equation}
and so appears with multiplicity four.  The
\begin{equation}
{\scriptstyle 1} \square_b
\end{equation}
sector in the effective $[X/{\mathbb Z}_2]$ orbifold
arises from any of the 
\begin{equation}
{\scriptstyle 1, a} \square_{b, ab}
\end{equation}
sectors; however, because of discrete torsion, the sectors
\begin{equation}
{\scriptstyle 1} \square_{b, ab}, \: \: \:
{\scriptstyle a} \square_{b, ab}
\end{equation}
contribute with opposite signs, and so cancel out of the one-loop
partition function in the $[X / {\mathbb Z}_2 \times {\mathbb Z}_2 ]$
orbifold.  Similarly, the 
\begin{equation}
{\scriptstyle b} \square_b, \: \: \:
{\scriptstyle ab} \square_{ab} \: \: \:
\mbox{ and } \: \: \:
{\scriptstyle ab} \square_b, \: \: \:
{\scriptstyle b} \square_{ab}
\end{equation}
one-loop sectors contribute with opposite signs because of discrete
torsion, and so cancel out.  As a result, the one-loop partition
function of $[X/ {\mathbb Z}_2 \times {\mathbb Z}_2]$ (with discrete
torsion) matches that
of a sigma model on (one copy of ) $X$, not a disjoint union.

Implicit in the analysis above, discrete torsion introduces an asymmetry
which breaks the ${\mathbb Z}_2^{(1)}$ one-form symmetry.  Recall that
this symmetry acts by tensoring existing bundles by 
${\mathbb Z}_2$ bundles.  Thus, in order for the symmetry to exist,
one would need, for example, the
\begin{equation}
{\scriptstyle 1} \square_{b, ab}
\end{equation}
sectors to enter the partition function symmetrically with the
\begin{equation}
{\scriptstyle a} \square_{b, ab}
\end{equation}
sectors, as they differ by tensor product with a ${\mathbb Z}_2$ bundle
given by
\begin{equation}
{\scriptstyle a} \square_1.
\end{equation}
Since discrete torsion weights those sectors differently,
the ${\mathbb Z}_2^{(1)}$ one-form symmetry is necessarily broken in this
model with discrete torsion in the noneffectively-acting `directions.'

This analysis trivially extends to the case of 
the orbifold $[X/ {\mathbb Z}_k \times {\mathbb Z}_k ]$,
where the first ${\mathbb Z}_k$ acts trivially and the second,
nontrivially.  (This example was discussed in
\cite[section 10.2]{Hellerman:2006zs}.)
Here, $H^2({\mathbb Z}_k \times {\mathbb Z}_k, U(1)) = 
{\mathbb Z}_k$, there are $k$ possible values for discrete torsion.

If one does not turn on discrete torsion at all, then all the twisted
sectors are symmetric with respect to one another, and tensoring in
${\mathbb Z}_k \times {\mathbb Z}_k$ bundles simply permutes them.
In this case, the theory admits a ${\mathbb Z}_k^{(1)}$
one-form symmetry.  In this case, with no discrete torsion, the theory
is equivalent to a sigma model on a disjoint union of
$k^2$ copies of $X$.

If we turn on discrete torsion, then the symmetry between the one-loop
sectors is broken, exactly as we saw in the ${\mathbb Z}_2 \times
{\mathbb Z}_2$ case, so there is no longer a 
${\mathbb Z}_k^{(1)}$ one-form symmetry, and hence the theory need not
decompose.  Indeed, computations in \cite[section 10.2]{Hellerman:2006zs}
indicated that in this example, for any nonzero value of discrete torsion,
the orbifold $[X/{\mathbb Z}_2 \times {\mathbb Z}_2]$ is equivalent
to a sigma model on one copy of $X$.

In terms of decomposition of two-dimensional theories with
one-form symmetries, turning on discrete torsion breaks the one-form
symmetry and so breaks the decomposition.  This is discussed in
\cite[section 10]{Hellerman:2006zs},
which discusses explicitly how decomposition is broken in cases where
the one-form symmetry is broken.

\subsection{$(-)^F$}

Another, more subtle, way to break naive one-form symmetries is
to modify a ${\mathbb Z}_2$ orbifold with a $(-)^F$ factor,
as implicit in the analysis of
\cite[section 2]{Hori:2011pd}. 
(See also \cite{Wong:2017cqs} for an excellent
recent review and overview of the application to
gauged linear sigma models \cite{Caldararu:2007tc}.)
We will see momentarily that the choice of orbifold can `make or break'
the existence of a ${\mathbb Z}_2^{(1)}$ one-form symmetry.

Consider for example a
${\mathbb Z}_2$ orbifold, acting on a two-dimensional (2,2)
supersymmetric theory with
matter.
There are two different ${\mathbb Z}_2$ orbifolds, one of which incorporates
an action of $(-)^F$.  Curiously, as discussed in \cite[section 2]{Hori:2011pd},
even if all of the matter is massive, the IR behavior is sensitive to the
orbifold, in a manner that depends upon the number of massive fields
(which are assumed to be acted upon nontrivially by the orbifold).

If all the matter on which the orbifold acts is massive,
then at low energies, below the scale of that matter,
one naively expects the theory to have a ${\mathbb Z}_2^{(1)}$ one-form
symmetry.  If that symmetry is indeed present, the decomposition predicts
the number of vacua should be even.  However, it was shown in
\cite[section 2]{Hori:2011pd} that whether the
number of vacua is even depends upon the number of massive fields upon which
the orbifold acts, as well as the type of orbifold and whether the
masses are complex masses or twisted masses.  In particular, if the
number of vacua is odd, then a ${\mathbb Z}_2^{(1)}$ one-form
symmetry cannot be present in the theory.

Briefly, in a ${\mathbb Z}_2$ orbifold in which the number of chirals
with complex masses, acted upon by the orbifold, is even,
or in a ${\mathbb Z}_2 (-)^F$ orbifold in which the number of chirals
with twisted masses, acted upon by the orbifold, is even,
the number of vacua is even, consistent with the existence of
a ${\mathbb Z}_2^{(1)}$ one-form symmetry.  In other cases,
one would not expect a one-form symmetry, based on the number of vacua.
(See for example
\cite{Gu:2019zkw} for a more detailed overview of the vacuum counting in
these theories.)

So far our analysis has been based solely on state counting.
We can also outline the obstruction at the level of orbifold twisted
sectors.
For example, multiplying in 
\begin{equation}
{\scriptstyle 1} \square_z
\end{equation}
maps between twisted and untwisted sector states, so, briefly,
existence of a one-form
symmetry implies a symmetry between the two, and from the
computations in \cite[section 2]{Hori:2011pd} 
that symmetry is only present in certain cases.

In passing, the even/odd distinction here is reminiscent of analogous
even/odd distinctions in existence and behavior of one-form symmetries
discussed in \cite{Cherman:2019hbq}.

\section{Gauging one-form symmetries in orbifolds}
\label{sect:gauging}

In this section we discuss gauging one-form symmetries in orbifolds,
setting up the technology we will use in specific examples in the
next section.  When we discuss specific examples, we will see that this
gauging has the effect of projecting onto particular summands in the
decomposition of the two-dimensional theory with a one-form symmetry.

In an ordinary orbifold $[X/H]$,
we sum over principal $H$-bundles on the worldsheet, each of which defines
a twisted sector of the orbifold.  For example, the partition function
for the case that the worldsheet is $T^2$ has the form
\begin{equation}
Z \: = \: \frac{1}{|H|} \sum_{gh = hg} {\scriptstyle g} \square_h,
\end{equation}
where we have used the notation
\begin{equation}
{\scriptstyle g} \square_h
\end{equation}
to indicate the contribution to the partition function from a sector with
boundary conditions / branch cuts determined by commuting element
$g, h \in H$.  Each such commuting pair $gh = hg$ defines a
principle $H$ bundle on the worldsheet, so the partition function sums
over principal $H$-bundles.

For an orbifold $[X/H]$ to have a $B G = G^{(1)}$ one-form symmetry 
means that 
$G \subset H$ is a subgroup of the orbifold group that acts trivially
on the theory
\cite{Pantev:2005zs,Pantev:2005rh,Pantev:2005wj}.  
Since $G$ acts trivially, for example, partition functions
are invariant under permutations defined by tensoring with
$G$ bundles, which defines the action of the one-form symmetry.
For example, if $G$ acts trivially on the theory, and
$z \in G$, then we can map
\begin{equation}
{\scriptstyle a} \square_b \: \mapsto \: {\scriptstyle a} \square_{z b},
\: \: \:
{\scriptstyle z a} \square_b,
\end{equation}
defined by the $G$ bundles
\begin{equation}
{\scriptstyle 1} \square_z, \: \: \:
{\scriptstyle z} \square_1,
\end{equation}
respectively.
This is a symmetry of the theory, since
the sectors which are exchanged make the same contribution to the
partition function -- because $G$ acts trivially.

In an orbifold $[X/H]$ with a one-form symmetry $B G = G^{(1)}$,
to gauge\footnote{
One might ask whether there can be gauge anomalies in gauged one-form
symmetries.  Certainly this can happen in theories in odd dimensions.
Consider for example an abelian Chern-Simons theories in three dimensions,
whose action can be recast as an integral of $F \wedge F$ over
a bounding four-manifold.  Gauging a one-form symmetry, schematically,
means replacing $F$ by $F-B$, resulting in a term involving
an integral of $B \wedge B$ over a bounding four-manifold, interpretable
in terms of an anomaly inflow to cancel what must be a one-form gauge
anomaly in three dimensions.  For a two-form $B$, one can only get
such interactions by integrating wedge powers of $B$ in even
dimensions, suggesting that one-form gauge anomalies only arise in
odd dimensions, and hence are not relevant for this paper, which
focuses on two-dimensional theories.  We would like to thank
Y.~Tanizaki and M.~\"Unsal for this observation.
} the one-form
symmetry means to 
\begin{enumerate}
\item sum over banded
$G$-gerbes on the worldsheet (i.e., $BG = G^{(1)}$ bundles), and
\item for each $G$-gerbe, sum over $H$ bundles which have
been twisted by the $G$ gerbe.
\end{enumerate}
Finally, the sum over gerbes can be weighted by a gerbe-dependent
phase factor $\epsilon(z)$, 
one phase for all contributions to that gerbe sector.
In a two-dimensional theory, this phase factor will take the form
\begin{equation}
\epsilon(z) \: \equiv \: \exp\left( i \langle \xi, z \rangle \right),
\end{equation}
where $z \in H^2(\Sigma, C^{\infty}(G))$ is the characteristic class
of the banded gerbe on the worldsheet $\Sigma$, and $\xi$ is a fixed
(gerbe-independent) class.

Let us take a moment to explain a few details of this procedure.
First, we have specified a sum over `banded' gerbes.  For a group $G$,
these are the $G$-gerbes on a space $M$ classified by
$H^2(M, C^{\infty}(G))$, and given on triple overlaps by
cocycles $h_{\alpha \beta \gamma}$.  If $G=U(1)$, these are the gerbes whose
connection is a two-form $B$ field -- and so summing over banded gerbes
naturally correlates with performing a path integral over $B$ fields.
However, banded gerbes are special cases of a more general class
of gerbes, which are also $BG = G^{(1)}$ bundles.  (See for
example \cite[section 3]{Hellerman:2006zs} for a more detailed discussion of the
differences.)  Connections on the
more general class involve both a $B$ field as well as an ordinary
gauge field on an associated outer automorphism bundle, and so summing
over nonbanded gerbes as well as banded gerbes would introduce an extra
field beyond what has been previously discussed in the literature.
In any event, we shall see explicitly in examples that restricting
the path integral to banded gerbes, instead of more general gerbes,
seems to be the physically consistent choice, and we will leave
more thorough treatments of why gauging one-form symmetries involves
path integrals that only sum over banded
gerbes instead of general gerbes for future work.

By `twisting the $H$ bundle,' we mean that the transition functions of
the $H$ bundle on the worldsheet only close on triple overlaps up to
the transition functions of the $G$ gerbe, i.e.
\begin{equation}
g_{\alpha \beta} g_{\beta \gamma} g_{\gamma \alpha} \: = \:
h_{\alpha \beta \gamma}
\end{equation}
for $g_{\alpha \beta}$ the transition functions on double overlaps
defining the $H$ bundle, and $h_{\alpha \beta \gamma}$ the transition
functions on triple overlaps defining the banded gerbe.
(This is the same twisting one encounters when describing Chan-Paton
factors in nontrivial $B$ field backgrounds; see for example
\cite{Anderson:2013sia} 
for a more detailed discussion of twisted bundles on gerbes.)
For $H$ bundles on a $T^2$, this means that instead of defining a bundle
by a commuting pair $g, h \in H$, we now work with almost-commuting pairs
$g, h$, where
\begin{equation}
g h \: = \: h g z
\end{equation} 
for $z \in G$.  We will denote contributions to partition functions
satisfying these boundary conditions by
\begin{equation}
{\scriptstyle g} \tsquare{z}_h.
\end{equation}

As a result, the partition function of a theory in which one gauges
a one-form action $B G = G^{(1)}$ on an orbifold $[X/H]$, for the case that the
worldsheet is $T^2$, takes the form
\begin{equation}
Z \: = \: \frac{1}{|G|} \frac{1}{|H|} \sum_{z \in G}
\sum_{g h = h g z} {\scriptstyle g} \tsquare{z}_h.
\end{equation}
(The sum over $z \in G$ is a sum over characteristic classes of banded $G$
gerbes, since for abelian $G$ on an oriented compact Riemann surface $\Sigma$,
$H^2(\Sigma, G) = G$.)
Furthermore, we can turn on an analogue of a discrete theta angle or
discrete torsion,
to weight gerbe-twisted sectors by phases $\epsilon(z)$:
\begin{equation}
Z \: = \: \frac{1}{|G|} \frac{1}{|H|} \sum_{z \in G} \epsilon(z)
\sum_{g h = h g z} {\scriptstyle g} \tsquare{z}_h.
\end{equation}
The phases $\epsilon(z)$ are the same phases described earlier -- they
determine different $G^{(1)}$ one-form group
actions on the theory, and in the case of orbifolds, are consistent with
modular invariance and factorization constraints as we shall
see later.
Also note that from the general expression given earlier,
$\epsilon(+1) = +1$.

In passing, note that in this fashion one can recover all elements
of the original group $H$:  for any $g, h \in H$,
$gh = hg z$ for $z = (hg)^{-1} (gh) = g^{-1} h^{-1} g h$.

Next, let us consider the action of the modular group, to check that the
expression above is well-defined, and to understand what constraints are
imposed by modular invariance.
$SL(2,{\mathbb Z})$ always acts
on $T^2$ twisted sectors with boundary conditions defined by
$(g,h)$ by mapping
\begin{equation}
(g,h) \: \mapsto \: (g^a h^b, g^c h^d).
\end{equation}

In an ordinary orbifold, this maps twisted sectors to twisted sectors:
since $g$ and $h$ commute, so too do $g^a h^b$ and
$g^c h^d$.
Here, consider a gerbe-twisted sector, twisted by $z \in G$.
Assume for the moment that $z$ lies in the center of $G$.
We will argue that, for $z$ in the center of $G$,
 $SL(2,{\mathbb Z})$ again maps
$z$-twisted sectors to $z$-twisted sectors.
It is straightforward to show that
\begin{equation}
g^a h^b \: = \: h^b g^a z^{ab},
\end{equation}
hence
\begin{equation}
g^a h^b g^c h^d \: = \: g^c h^d g^a h^b z^{ad-bc}
\: = \:  g^c h^d g^a h^b z,
\end{equation}
(since for an $SL(2,{\mathbb Z})$ matrix, $ad-bc=1$,)
hence we propose that modular transformations map
\begin{equation}
{\scriptstyle g} \tsquare{z}_h
\end{equation}
to
\begin{equation}
{\scriptstyle g^a h^b} \tsquare{ z }_{g^c h^d},
\end{equation}
remaining within the $z$-gerbe-twisted sector.

In general, for $z$ not necessarily in the center,
$SL(2,{\mathbb Z})$ transformations will exchange different
gerbe characteristic classes.
Consider for example the $SL(2,{\mathbb Z})$ transformation
\begin{equation}
\left[ \begin{array}{cc}
1 & 1 \\ 0 & 1 \end{array} \right].
\end{equation}
This modular transformation maps the pair $(g,h) \mapsto (gh, h)$.
Now, if $gh = hg z$, then
\begin{equation}
gh h \: = \: hg z h \: = \:  h g h ( h^{-1} z h),
\end{equation}
and so we see that the original $z$-gerbe sector is mapped to
an $h^{-1} z h$-gerbe sector.  (We will see a more detailed example
of this form in section~\ref{sect:ex4}.)

As a quick consistency check, let us quickly note that $h^{-1} z h$ is still in
the group defining the gerbe.  Suppose $K$ is any normal subgroup of $H$:
\begin{equation}
1 \: \longrightarrow \: K \: \longrightarrow \: H \: 
\stackrel{f}{\longrightarrow} \: G \: \longrightarrow \: 1.
\end{equation}
Let $z \in K$.  This implies that $f(z) = 1$.
Now,
\begin{eqnarray}
f( h^{-1} z h ) & = & f(h)^{-1} f(z) f(h),
\\
& = & f(h)^{-1} f(h),
\\
& = & 1,
\end{eqnarray}
and so we see that $h^{-1} z h \in K$ also.

As a result, in these theories,
modular invariance requires us to sum over all gerbes
(in addition to bundles), in the same fashion that in ordinary
orbifolds, modular invariance requires us to sum over all twisted
sectors defined by bundles.

Now, let us turn to the phases $\epsilon(z)$ and constraints imposed
by modular invariance.  In general, modular transformations map $z$ to some
conjugacy class within the orbifold group, so $\epsilon(z)$ must be constant
on those conjugacy classes -- it must be the restriction of a class function
on the orbifold group to the gerbe group $G$, in other words.
If for example the orbifold group is abelian, then the set of
$z$-gerbe twisted sectors close into themselves under $SL(2,{\mathbb Z})$,
and so no constraints are imposed upon the phases $\epsilon(z)$.
(In fact, if the orbifold group is abelian, then the only sectors that
arise in a partition function are those for which $z=1$, so $\epsilon(z)$
for $z \neq 1$ is irrelevant.)

Modular invariance is not the only criterion that orbifold phase factors
must obey; they must also be consistent with factorization
(see e.g. \cite[section 4.3]{Sharpe:2000ki}).  
This says that if a multiloop diagram
can degenerate into a product of diagrams with fewer loops, separated
by long thin tubes, then since phase factors are moduli independent,
the phase factor associated to the multiloop diagram should be the
product of the phase factors associated to the diagrams with fewer loops.

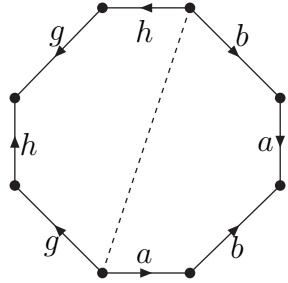
\begin{figure}[ht]
\centering
\begin{picture}(100,100)(0,0)
\ArrowLine(33,0)(66,0)
\ArrowLine(66,0)(100,33)
\ArrowLine(100,66)(100,33)
\ArrowLine(66,100)(100,66)
\ArrowLine(33,0)(0,33)
\ArrowLine(0,33)(0,66)
\ArrowLine(33,100)(0,66)
\ArrowLine(66,100)(33,100)
\Vertex(33,0){2}
\Vertex(66,0){2}
\Vertex(100,33){2}
\Vertex(100,66){2}
\Vertex(66,100){2}
\Vertex(0,33){2}
\Vertex(0,66){2}
\Vertex(33,100){2}
\Text(49,4)[b]{$a$}
\Text(84,14)[t]{$b$}
\Text(14,14)[t]{$g$}
\Text(2,49)[l]{$h$}
\Text(49,96)[t]{$h$}
\Text(86,86)[b]{$b$}
\Text(16,86)[b]{$g$}
\Text(98,49)[r]{$a$}
\DashLine(33,0)(66,100){2}
\end{picture}
\caption{A schematic illustration of an orbifold two-loop diagram,
i.e. a contribution to an orbifold partition function on a genus two 
surface.  We have assumed that the group elements are such that the
diagram factorizes into a product of two one-loop diagrams, in which
effectively the dashed line shrinks to a point. \label{fig:2loop}}
\end{figure}

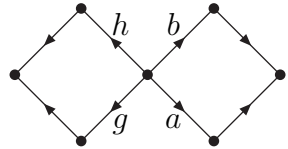
\begin{figure}[ht]
\centering
\begin{picture}(100,50)(0,0)
\ArrowLine(50,25)(25,0)
\ArrowLine(50,25)(25,50)
\ArrowLine(25,0)(0,25)
\ArrowLine(25,50)(0,25)
\ArrowLine(50,25)(75,0)
\ArrowLine(50,25)(75,50)
\ArrowLine(75,0)(100,25)
\ArrowLine(75,50)(100,25)
\Vertex(0,25){2}
\Vertex(25,0){2}
\Vertex(25,50){2}
\Vertex(50,25){2}
\Vertex(75,0){2}
\Vertex(75,50){2}
\Vertex(100,25){2}
\Text(40,10)[t]{$g$}
\Text(40,40)[b]{$h$}
\Text(60,10)[t]{$a$}
\Text(60,40)[b]{$b$}
\end{picture}
\caption{A two-loop diagram, factored into a pair of one-loop diagram.
\label{fig:fact}}
\end{figure}

In figure~\ref{fig:2loop}, we have schematically illustrated an
orbifold two-loop diagram, and figure~\ref{fig:fact} we have illustrated
its factorization.

If the two one-loop diagram factors are in sectors twisted by gerbe
characteristic classes $z_1$, $z_2$, then, in the case that $G$ lies in
the center of $H$, the two-loop diagram twisted by gerbe characteristic
class $z_1 z_2$.  In two dimensional theories on a space $\Sigma$, 
since $H^2(\Sigma)$
is one-dimensional, independent of the genus of $\Sigma$, the phases
associated to two-loop diagrams are the same as the phases associated to
one-loop diagrams, and so we find, again for the case that $G$ is in the
center of the orbifold group, that
\begin{equation}
\epsilon(z_1 z_2) \: = \: \epsilon(z_1) \epsilon(z_2).
\end{equation}
Thus, formally\footnote{Not every gerbe twisted sector might be
nonempty, so for some values of $z$, there may be no corresponding
twisted sectors.  As a result, 
this argument should be interpreted as giving a consistency
test on existing results.},
factorization implies that $\epsilon: G \rightarrow U(1)$ is
a group homomorphism, which is consistent with its first-principles
description, given earlier, as a phase factor derived from the characteristic
class of the banded gerbe on the worldsheet.

We leave a discussion of factorization in non-banded gerbes,
in which $G$ does not lie in the center of $H$, for future work.

So far we have discussed partition functions.  Next, we shall briefly
turn our attention to massless spectra.  Of course, once one has computed
the partition function, one can read off the massless spectrum, so a 
separate discussion of massless spectra in orbifolds in which a
one-form symmetry has been gauged is somewhat redundant. As a result,
we will be brief.  In a nutshell, because the one-form symmetry
permutes twisted sectors, e.g. schematically
\begin{equation}
{\scriptstyle a} \square_b \: \mapsto \:
{\scriptstyle z a} \square_b, \: \: \:
{\scriptstyle a} \square_{z b},
\end{equation}
the effect of the one-form-symmetry-gauging in massless spectra is very
closely related to massless spectra in orbifolds in which one gauges
a quantum symmetry, as in e.g. \cite[section 8.3]{Ginsparg:1988ui}.
We note that if one gauges the ${\mathbb Z}_k$ quantum symmetry of
a ${\mathbb Z}_k$ orbifold, one recovers the original unorbifolded theory,
which is closely related to the observation of this work that gauging
one-form symmetries selects out summands of decomposition.
In any event, since massless spectra can be read off from partition functions,
we shall focus on partition functions in our discussions of orbifolds.

\section{Examples in orbifolds}
\label{sect:exs-orbs}

In this section we will describe apply the technology set up in the
previous section to particular examples of orbifolds with global
one-form symmetries.
We will see that gauging the one-form symmetry has the effect of projecting
onto particular summands in the decomposition of the original theory.

\subsection{Trivial ${\mathbb Z}_k$ gerbe}

Let us begin with the simplest possible example, the orbifold
$[X / {\mathbb Z}_k]$ where all of ${\mathbb Z}_k$ acts trivially
on $X$.  (The details of $X$ are irrelevant; we assume it is Calabi-Yau
so that this theory is a CFT, for simplicity.)  This theory has a 
global ${\mathbb Z}_k^{(1)} =  B {\mathbb Z}_k$ one-form symmetry,
which on the partition function 
\begin{equation}
Z \: = \: \frac{1}{| {\mathbb Z}_k | } \sum_{g, h} {\scriptstyle g} \square_h
\end{equation}
simply permutes the various twisted
sectors.  As discussed in \cite[section 5.1]{Hellerman:2006zs},
decomposition predicts that this theory is equivalent to a disjoint
union of $k$ copies of $X$, one copy for each irreducible representation
of ${\mathbb Z}_k$.  Furthermore, since the ${\mathbb Z}_k$ gerbe is trivial,
there is no difference in $B$ fields -- each copy of $X$ in the decomposition
has the same $B$ field.

Now, we shall gauge the ${\mathbb Z}_k^{(1)} = B {\mathbb Z}_k$ 
one-form symmetry.  Including possible phases $\epsilon(z)$,
the partition function of the new theory has the form
\begin{equation}
Z \: = \:
\frac{1}{|{\mathbb Z}_k|} \frac{1}{|{\mathbb Z}_k|}
\sum_{z \in {\mathbb Z}_k} 
\epsilon(z)
\sum_{g h = h g z} {\scriptstyle g} \tsquare{z}_h.
\end{equation}
However, in this theory, only when $z=1$ are there any solutions to 
$g h = h g z$.  If $z \neq 1$, then there are no $g, h$ satisfying that
equation, because all elements of the group commute.
As a result, using $\epsilon(1) = 1$,
the partition function simplifies to
\begin{equation}
Z \: = \:
\frac{1}{|{\mathbb Z}_k|} \frac{1}{|{\mathbb Z}_k|}
\sum_{gh = hg} {\scriptstyle g} \square_h,
\end{equation}
and since ${\mathbb Z}_k$ acts trivially,
\begin{equation}
{\scriptstyle g} \square_h \: = \: {\scriptstyle 1} \square_1 \: = \: Z(X),
\end{equation}
hence
\begin{equation}
Z \: = \:
\frac{1}{|{\mathbb Z}_k|} \frac{1}{|{\mathbb Z}_k|}
(k^2) 
Z(X) \: = \: Z(X).
\end{equation}

Thus, we see that gauging the one-form symmetry in the trivially-acting
orbifold returns the partition function of $X$, as expected.

\subsection{First nontrivial banded example}
\label{sect:1stbanded}

In this section, we will consider the first banded example
discussed in \cite[section 5.2]{Hellerman:2006zs}.
This is an orbifold $[X/D_4]$, where the ${\mathbb Z}_2$ center of
the eight-element group $D_4$ acts trivially.  As before, the details
of $X$ are not important; we assume $X$ is Calabi-Yau and admits an
effective action of ${\mathbb Z}_2 \times {\mathbb Z}_2$.

Now, $D_4 / {\mathbb Z}_2 = {\mathbb Z}_2 \times {\mathbb Z}_2$, or more
elegantly,
\begin{equation}
1 \: \longrightarrow \: {\mathbb Z}_2 \: \longrightarrow \: D_4 \:
\longrightarrow \: {\mathbb Z}_2 \times {\mathbb Z}_2 \: \longrightarrow \: 1.
\end{equation}
The elements of $D_4$ are
\begin{equation}
\{ 1, z, a, b, a z, b z, a b, b a = a b z \},
\end{equation}
where $z$ generates the ${\mathbb Z}_2$ center, $a^1 = 1$, $b^2 = z$.
We denote elements of the coset ${\mathbb Z}_2 \times {\mathbb Z}_2$ as
\begin{equation}
\{ \overline{1}, \overline{a}, \overline{b}, \overline{ab} \},
\end{equation}
where $\overline{1} = \{1, z\}$, $\overline{a} = \{a, az\}$, and so forth.

In \cite{Hellerman:2006zs}, it was argued that this orbifold decomposed
into a disjoint union of two copies of $[X/{\mathbb Z}_2 \times
{\mathbb Z}_2]$, one copy without discrete torsion, one copy with.
(In this example, \cite{Hellerman:2006zs} also computed multiloop partition
functions and massless spectra to double-check the claimed decomposition.)

Because the ${\mathbb Z}_2$ acts trivially,
the orbifold $[X / D_4]$ is a ${\mathbb Z}_2$ gerbe, which is to say,
it has a global ${\mathbb Z}_2$ one-form symmetry, also denoted
${\mathbb Z}_2^{(1)} = B {\mathbb Z}_2$.

We can gauge that ${\mathbb Z}_2$ one-form symmetry following the
procedure described above.  At one-loop, the partition function
has the form
\begin{equation}
Z \: = \: \frac{1}{|{\mathbb Z}_2|} \frac{1}{|D_4|} 
\sum_{z \in {\mathbb Z}_2} \sum_{gh = hgz} \epsilon(z) \,
{\scriptstyle g} \tsquare{z}_h.
\end{equation}

Now, for the trivial gerbe, when $z=1$, the allowed one-loop twisted
sectors (defined by commuting $g$, $h$) are
\begin{equation}
{\scriptstyle 1, z} \square_{1, z}, \: \: \:
{\scriptstyle 1, z} \square_{a, az}, \: \: \:
{\scriptstyle 1, z} \square_{b, bz}, \: \: \:
{\scriptstyle 1, z} \square_{ab, ba}, \: \: \:
{\scriptstyle a, az} \square_{a, az}, \: \: \:
{\scriptstyle b, bz} \square_{b, bz}, \: \: \:
{\scriptstyle ab, ba} \square_{ab, ba}.
\end{equation}
As $D_4$ is nonabelian, only commuting pairs are in the list above.
Projecting to the corresponding ${\mathbb Z}_2 \times {\mathbb Z}_2$
orbifolds, this list omits
\begin{equation}
{\scriptstyle \overline{a}} \square_{\overline{b}}, \: \: \:
{\scriptstyle \overline{a}} \square_{\overline{ab}}, \: \: \:
{\scriptstyle \overline{b}} \square_{\overline{ab}}
\end{equation}
twisted sectors, as these do not lift to commuting elements of $D_4$.
In the language of decomposition, each of these
${\mathbb Z}_2 \times {\mathbb Z}_2$ orbifolds is also multiplied by 
a sign when discrete torsion is turned on, so adding theories with
and without discrete torsion effectively cancels these out of partition
functions.

Next, for the nontrivial gerbe, when $z \neq 1$,
the only possible sectors are
\begin{equation}
{\scriptstyle a, az} \tsquare{z}_{b, bz}, \: \: \:
{\scriptstyle a, az} \tsquare{z}_{ab, ba}, \: \: \:
{\scriptstyle b, bz} \tsquare{z}_{ab, ba}.
\end{equation}
Note that when we project to ${\mathbb Z}_2 \times {\mathbb Z}_2$
orbifolds, these fill in precisely the twisted sectors that were missing
from the $D_4$ orbifold.

Putting this together, the one-loop partition function of the
theory after gauging $B {\mathbb Z}_2$ is
\begin{eqnarray}
Z & = & \frac{1}{| {\mathbb Z}_2 | } \frac{1}{| D_4 |} 
\Biggl[  {\scriptstyle 1, z} \square_{1, z} \: + \:
{\scriptstyle 1, z} \square_{a, az} \: + \:
{\scriptstyle 1, z} \square_{b, bz} \: + \:
{\scriptstyle 1, z} \square_{ab, ba} \: + \:
{\scriptstyle a, az} \square_{a, az} \: + \:
{\scriptstyle b, bz} \square_{b, bz} \: + \:
{\scriptstyle ab, ba} \square_{ab, ba}
\nonumber \\
& & \hspace*{0.75in}
\: + \: \epsilon(z) \left(
{\scriptstyle a, az} \tsquare{z}_{b, bz} \: + \:
{\scriptstyle a, az} \tsquare{z}_{ab, ba} \: + \:
{\scriptstyle b, bz} \tsquare{z}_{ab, ba}
\right) \Biggr],
\end{eqnarray}
and since $z$ acts trivially, we can reduce this to
${\mathbb Z}_2 \times {\mathbb Z}_2$ twisted sectors as
\begin{eqnarray}
Z & = & \frac{1}{4} 
\Biggl[
{\scriptstyle \overline{1}} \square_{ \overline{1}} \: + \:
{\scriptstyle \overline{1}} \square_{ \overline{a}} \: + \:
{\scriptstyle \overline{1}} \square_{ \overline{b}} \: + \:
{\scriptstyle \overline{1}} \square_{\overline{ab}} \: + \:
{\scriptstyle \overline{a}} \square_{\overline{a}} \: + \:
{\scriptstyle \overline{b}} \square_{\overline{b}} \: + \:
{\scriptstyle \overline{ab}} \square_{\overline{ab}}
\: + \: \cdots
\nonumber \\
& & \hspace*{0.75in}
\: + \: \epsilon(z) \left(
{\scriptstyle \overline{a}} \square_{\overline{b}} \: + \:
{\scriptstyle \overline{a}} \square_{\overline{ab}} \: + \:
{\scriptstyle \overline{b}} \square_{\overline{ab}} \right) \Biggr].
\end{eqnarray}

When $\epsilon(z) = + 1$, this is the one-loop partition function
of $[X/ {\mathbb Z}_2 \times {\mathbb Z}_2]$ without discrete torsion,
one factor in the decomposition,
and when $\epsilon(z) = -1$, this is the one-loop partition function
of $[X/ {\mathbb Z}_2 \times {\mathbb Z}_2]$ with discrete torsion,
the other factor in the decomposition.

A closely related example is discussed in
\cite[section 5.3]{Hellerman:2006zs}.
That example is a $[X/{\mathbb H}]$ orbifold,
where ${\mathbb H}$ is the eight-element group of quaternions,
and the ${\mathbb Z}_2$ center is taken to act trivially on $X$.
Just as in the $D_4$ example above, the theory decomposes into a disjoint
union of two $[X/{\mathbb Z}_2 \times {\mathbb Z}_2]$ orbifolds, one with
discrete torsion, and one without.  The analysis of both the
original theory as well as the gauging of the $B {\mathbb Z}_2$ symmetry
runs very closely parallel to the $D_4$ example just discussed, so we do not
repeat the details here.

\subsection{A non-banded example}

Next we consider an example discussed in
\cite[section 5.4]{Hellerman:2006zs}.  Specifically,
this is a $[X/{\mathbb H}]$ orbifold, where ${\mathbb H}$ is
the eight-element group of quaternions
\begin{equation}
\{ \pm 1, \pm i, \pm j, \pm k \},
\end{equation}
and the non-central subgroup $\langle i \rangle = \{ \pm 1, \pm i \}
\cong {\mathbb Z}_4$
acts trivially on $X$.  In this case, ${\mathbb H}/\langle i \rangle
= {\mathbb Z}_2$.  This is a non-banded example, and for the most part
in this paper we focus on banded examples, but we include this and
other non-banded examples to illustrate further complexities that arise.

The decomposition conjecture \cite{Hellerman:2006zs} predicts that
this orbifold decomposes into three disjoint components as
\begin{equation}
[ X / {\mathbb Z}_2] \coprod [X/{\mathbb Z}_2] \coprod X,
\end{equation}
which was checked in \cite[section 5.4]{Hellerman:2006zs} by 
computing one-loop and two-loop partition functions, and by
examining operators in the theory.

This theory has a $B {\mathbb Z}_4$ symmetry, which at the level of
the partition function corresponds to interchanging the twisted sectors.
Let us consider gauging that $B {\mathbb Z}_4$ symmetry.
Following the procedure described earlier, the one-loop partition function
of the new theory, the orbifold with the gauged $B {\mathbb Z}_4$, has
the form
\begin{equation}
Z \: = \: \frac{1}{| {\mathbb Z}_4|} \frac{1}{| {\mathbb H}|} 
\sum_{z \in {\mathbb Z}_4} \sum_{gh = hgz} 
\epsilon(z) \,
{\scriptstyle g} \tsquare{z}_h,
\end{equation}
where the $\epsilon(z)$ are possible phases.

For the $z=1$ gerbe, there are the following contributions:
\begin{equation}
{\scriptstyle \pm 1, \pm i} \square_{\pm 1, \pm i}, \: \: \:
{\scriptstyle \pm 1} \square_{\pm j, \pm k}, \: \: \:
{\scriptstyle \pm j, \pm k} \square_{\pm 1}, \: \: \:
{\scriptstyle \pm j} \square_{\pm j}, \: \: \:
{\scriptstyle \pm k} \square_{\pm k}.
\end{equation}
The only contribution, the only twisted bundles, in sectors with a nontrivial
gerbe are in the case $z = -1$:
\begin{equation}
{\scriptstyle \pm i} \tsquare{-1}_{\pm j, \pm k}, \: \: \:
{\scriptstyle \pm j, \pm k} \tsquare{-1}_{\pm i}, \: \: \:
{\scriptstyle \pm j} \tsquare{-1}_{\pm k}, \: \: \:
{\scriptstyle \pm k} \tsquare{-1}_{\pm j}.
\end{equation}

Putting this together, and writing this in terms of the twisted
sectors of an effectively-acting ${\mathbb Z}_2$ orbifold,
with group elements $\{1, \xi \}$, $\xi^2 = 1$, we have that
the one-loop partition function is
\begin{eqnarray}
Z & = & \frac{1}{| {\mathbb Z}_4|} \frac{1}{| {\mathbb H}|} 
 \Biggl[ 
16 \, {\scriptstyle 1} \square_1 \: + \:
8 \, {\scriptstyle 1} \square_{\xi} \: + \:
8 \, {\scriptstyle \xi} \square_1 \: + \:
4 \, {\scriptstyle \xi} \square_{\xi} \: + \:
4 \, {\scriptstyle \xi} \square_{\xi}
\nonumber \\
& & \hspace*{1.0in}
\: + \: \epsilon(-1)\left(
8 \, {\scriptstyle 1} \square_{\xi} \: + \:
8 \, {\scriptstyle \xi} \square_1 \: + \:
4 \, {\scriptstyle \xi} \square_{\xi} \: + \:
4 \, {\scriptstyle \xi} \square_{\xi} \right) \Biggr].
\end{eqnarray}

In the case that $\epsilon(-1) = +1$,
\begin{eqnarray}
Z & = & \frac{1}{4} \frac{16}{8} \left(
{\scriptstyle 1} \square_1 \: + \:
 {\scriptstyle 1} \square_{\xi} \: + \:
 {\scriptstyle \xi} \square_1 \: + \:
{\scriptstyle \xi} \square_{\xi} \right),
\\
& = & \frac{1}{2} \left(
{\scriptstyle 1} \square_1 \: + \:
 {\scriptstyle 1} \square_{\xi} \: + \:
 {\scriptstyle \xi} \square_1 \: + \:
{\scriptstyle \xi} \square_{\xi} \right)
\: = \: Z( [X/{\mathbb Z}_2] ),
\end{eqnarray}
the one-loop partition function of a ${\mathbb Z}_2$ orbifold.

In the case that $\epsilon(-1) = -1$,
\begin{equation}
Z \: = \: \frac{1}{4} \frac{16}{8} \left(
{\scriptstyle 1} \square_1 \right) \: = \: (1/2) Z(X),
\end{equation}
proportional to the one-loop partition function of $X$ itself.

\subsection{Another non-banded example}
\label{sect:ex4}

Next, consider the orbifold $[X/A_4]$, where $A_4$ is the 
twelve-element nonabelian
group of alternating permutations of four elements.  
This group has a subgroup ${\mathbb Z}_2 \times {\mathbb Z}_2$, which
we will take to act trivially.  It can be shown that
$A_4 / {\mathbb Z}_2 \times {\mathbb Z}_2 = {\mathbb Z}_3$.
Explicitly, in terms of permutations, the ${\mathbb Z}_2 \times
{\mathbb Z}_2$ normal subgroup is generated by
\begin{equation}
\alpha \: \equiv \: (14)(23), \: \: \:
\beta \: \equiv \: (13)(24), \: \: \:
\gamma \: \equiv \: (12)(34),
\end{equation}
and elements of the ${\mathbb Z}_3$ cosets are given by
\begin{equation}
\{ 1, \alpha, \beta, \gamma \}, \: \: \:
\{ (123), (142), (243), (134) \}, \: \: \:
\{ (132), (143), (124), (234) \}.
\end{equation}

This example was considered in \cite[section 5.5]{Hellerman:2006zs},
\cite[section 2.0.5]{Pantev:2005rh}
where it was shown that this theory decomposes into
\begin{equation}
[X/{\mathbb Z}_3] \coprod X.
\end{equation}

This theory has a $B( {\mathbb Z}_2 \times {\mathbb Z}_2)$ symmetry,
which we will gauge.
Proceeding as before, the one-loop partition function of the
orbifold with $B( {\mathbb Z}_2 \times {\mathbb Z}_2)$ gauged has the
form
\begin{equation}  \label{eq:z-2ndnonband}
Z \: = \: \frac{1}{| {\mathbb Z}_2 \times {\mathbb Z}_2|}
\frac{1}{|A_4|} \sum_{z \in {\mathbb Z}_2 \times {\mathbb Z}_2}
\sum_{gh = hgz} \epsilon(z) \,
{\scriptstyle g} \tsquare{z}_h.
\end{equation}

Let $\xi$ denote the generator of ${\mathbb Z}_3$.  As noted in
\cite[section 2.0.5]{Pantev:2005rh},
the
\begin{equation}
{\scriptstyle 1} \square_1
\end{equation}
sector of the ${\mathbb Z}_3$ orbifold arises from 16 sectors of the
$[X/A_4]$ orbifold (with trivial gerbe), and the other
${\mathbb Z}_3$ twisted sectors arise from only 4 sectors of the
$[X/A_4]$ orbifold (with trivial gerbe).
For example, the
\begin{equation}
{\scriptstyle 1} \square_{\xi}
\end{equation}
sector of the ${\mathbb Z}_3$ orbifold arises from the
\begin{equation}
{\scriptstyle 1} \square_{(123)}, \: \: \:
{\scriptstyle 1} \square_{(142)}, \: \: \:
{\scriptstyle 1} \square_{(243)}, \: \: \:
{\scriptstyle 1} \square_{(134)}
\end{equation}
one-loop twisted sectorss of the $[X/A_4]$ orbifold.

Related contributions arise in gerbe-twisted sectors.
For example, the $\gamma$-twisted sectors include
\begin{equation}
{\scriptstyle \alpha} \tsquare{$\gamma$}_{(123)}, \: \: \:
{\scriptstyle \alpha} \tsquare{$\gamma$}_{(142)}, \: \: \:
{\scriptstyle \alpha} \tsquare{$\gamma$}_{(243)}, \: \: \:
{\scriptstyle \alpha} \tsquare{$\gamma$}_{(134)}, 
\end{equation}
which also project to the 
\begin{equation}
{\scriptstyle 1} \square_{\xi}
\end{equation}
sector of the ${\mathbb Z}_3$ orbifold.

In this example, the action of $SL(2,{\mathbb Z})$ will rotate between
different gerbes.  To see this, first note that
\begin{equation}
\alpha (123) \: = \: (123) \alpha \gamma,
\end{equation}
as can be seen explicitly as follows:
\begin{eqnarray}
\alpha (123): & 1234 & \stackrel{(123)}{\longrightarrow} \:
2314 \: \stackrel{\alpha} \: 3241,
\\
(123) \alpha: & 1234 & \stackrel{\alpha}{\longrightarrow} \:
4321 \: \stackrel{(123)} \: 4132,
\end{eqnarray}
and the two outputs are related by $\gamma$.
As a result,
\begin{equation}
(123)^{-1} \alpha (123) \: = \: \alpha \gamma \: = \: \beta.
\end{equation}
In addition:
\begin{eqnarray}
(123)^{-1} \beta (123) & = & \beta \alpha \: = \: \gamma,
\\
(123)^{-1} \gamma (123) & = & \gamma \beta \: = \: \alpha,
\end{eqnarray}
From the analysis of section~\ref{sect:gauging}, we see therefore that
modular transformations will relate all gerbes of characteristic
class $z \neq 1$, in addition to
twisted sectors.

As a result, to add phases $\epsilon(z)$ consistent with modular
invariance, $\epsilon$ must be constant on $z \neq 1$.

Finally, we can write the partition function~(\ref{eq:z-2ndnonband}) in terms
of ${\mathbb Z}_3$ orbifold twisted sectors as follows:
\begin{eqnarray}
Z & = & 
 \frac{1}{| {\mathbb Z}_2 \times {\mathbb Z}_2|}
\frac{1}{|A_4|} 
\left( 16 \, {\scriptstyle 1} \square_1 \: + \:
4 (1 + 3 \epsilon) \, {\scriptstyle 1} \square_{\xi} \: + \:
4 (1 + 3 \epsilon) \, {\scriptstyle 1} \square_{\xi^2} \: + \:
4 (1 + 3 \epsilon) \, {\scriptstyle \xi} \square_{\xi^2} \: + \:
\cdots \right),
\nonumber \\
\end{eqnarray}
where we have taken $\epsilon(1) = 1$ and $\epsilon(z) = \epsilon$
for $z \neq 1$.

In the case $\epsilon = +1$, this can be simplified to
\begin{eqnarray}
Z & = & \frac{1}{3} \left( {\scriptstyle 1} \square_1
\: + \:
 {\scriptstyle 1} \square_{\xi} \: + \:
{\scriptstyle 1} \square_{\xi^2} \: + \:
{\scriptstyle \xi} \square_{\xi^2} \: + \:
\cdots \right),
\end{eqnarray}
which is the partition function of $[X/{\mathbb Z}_3]$.

If $\epsilon = -1/3$, the partition function is proportional to
that of $X$, the other component of the decomposition.

\subsection{A nonabelian gerbe}

Consider the orbifold $[X/D_4]$, where now all of $D_4$ acts trivially
on $X$.  This is a $D_4$ gerbe over $X$, a generalized space with
fibers $B D_4$.  However, the one-form symmetry is determined by the
center of $D_4$, which is ${\mathbb Z}_2$.  The two-group
$B {\mathbb Z}_2$ acts on $B D_4$, and acts on $[X / D_4]$.

From \cite{Hellerman:2006zs}, the decomposition of this theory has
as many components as irreducible representation of $D_4$, of which
there are five:  four one-dimensional representations and one two-dimensional
representation.  This theory is therefore equivalent to a sigma model
on a disjoint union of five copies of $X$.

Now, let us consider gauging the $B {\mathbb Z}_2 = {\mathbb Z}_2^{(1)}$ 
one-form symmetry of
this orbifold.  
Since we can only\footnote{
$D_4$ is nonabelian, but one-form symmetries correlate to
abelian groups.  Hence, there is an action of a center one-form
symmetry (here, ${\mathbb Z}_2^{(1)}$) on the $D_4$ gerbe,
but we cannot define a $D_4^{(1)}$ symmetry.}
gauge
the action of ${\mathbb Z}_2^{(1)}$ on the $D_4$ gerbe, we should not
expect gauging alone to select out each individual component, but at
best merely groups of the five components of the decomposition,
and indeed that is what we shall find.

Proceeding as before, the partition function on $T^2$
takes the form
\begin{equation}
Z \: = \: \frac{1}{| {\mathbb Z}_2|} \frac{1}{|D_4|}
\sum_{z \in {\mathbb Z}_2} \sum_{g h = h g z}
\epsilon(z) \, 
{\scriptstyle g} \tsquare{z}_h.
\end{equation}
Proceeding as in section~\ref{sect:1stbanded},
the one-loop partition function
of the theory after gauging $B {\mathbb Z}_2$ is
\begin{eqnarray}
Z & = & \frac{1}{| {\mathbb Z}_2 | } \frac{1}{| D_4 |} 
\Biggl[  {\scriptstyle 1, z} \square_{1, z} \: + \:
{\scriptstyle 1, z} \square_{a, az} \: + \:
{\scriptstyle 1, z} \square_{b, bz} \: + \:
{\scriptstyle 1, z} \square_{ab, ba} \: + \:
{\scriptstyle a, az} \square_{a, az} \: + \:
{\scriptstyle b, bz} \square_{b, bz} \: + \:
{\scriptstyle ab, ba} \square_{ab, ba}
\nonumber \\
& & \hspace*{0.75in} 
\: + \: \epsilon(z) \left(
{\scriptstyle a, az} \tsquare{z}_{b, bz} \: + \:
{\scriptstyle a, az} \tsquare{z}_{ab, ba} \: + \:
{\scriptstyle b, bz} \tsquare{z}_{ab, ba}
\right) \Biggr],
\end{eqnarray}
and since all group elements act trivially, this reduces to
\begin{equation}
Z \: = \: \frac{1}{16} \left( 40 + 24 \epsilon(-1) \right)
{\scriptstyle 1} \square_1.
\end{equation}

When $\epsilon(-1) = +1$,
\begin{equation}
Z \: = \: 4 Z(X),
\end{equation}
and when $\epsilon(-1) = -1$,
\begin{equation}
Z \: = \: Z(X).
\end{equation}
It appears that one choice of phase $\epsilon$ selects
four copies of $X$, and the other choice selects one copy of
$X$, for a total of five copies, reproducing the decomposition of
$[X/D_4]$, and also mimicking the structure of the irreducible
representation of $D_4$:  four irreducible representations of
dimension one, and one of dimension two.

\subsection{Dijkgraaf-Witten theory in two dimensions}

Two-dimensional Dijkgraaf-Witten theory \cite{Dijkgraaf:1989pz} is
defined by a two-dimensional quantum field theory which is essentially
an orbifold of a point.  The theory $[{\rm point}/G]$ clearly is associated
with a gerbe, and is essentially a degenerate limit of the orbifolds
discussed previously in this section.  It can be twisted by turning
on discrete torsion, but as we observed in section~\ref{sect:dt},
turning on discrete torsion typically breaks two-group symmetries,
so we will only consider theories without discrete torsion. 

As our analysis amounts to an application of ideas discussed extensively
elsewhere in this paper, we shall be brief.
Following decomposition \cite{Hellerman:2006zs},
the two-dimensional orbifold $[{\rm point}/G]$ ($G$ finite)
is equivalent
to a sigma model on as many points as irreducible representations of
$G$.  If $G$ is abelian, the theory has a $BG = G^{(1)}$ one-form symmetry,
and gauging that one-form symmetry can be used to select particular
summands in that decomposition, as has been discussed elsewhere in this
paper.  If $G$ is nonabelian, then only a $BK = K^{(1)}$ one-form
symmetry is manifest, for $K$ the center of $G$; however, if the number
of irreducible representations of $G$ is greater than $|K|$, then the
theory will have additional, less obvious, one-form symmetries which can be
used to select each summand in the decomposition.

\section{Nonsupersymmetric pure Yang-Mills theories in two dimensions}
\label{sect:pure-ym}

So far, we have discussed orbifolds with one-form symmetries,
and the gauging of those one-form symmetries.  In this and
the next several
sections, we turn our attention to two-dimensional
gauge theories, and perform similar
analyses.  Specifically, we will describe several families of 
two-dimensional gauge theories with one-form symmetries and their
gauging.  Just as in the case of orbifolds, decomposition
\cite{Hellerman:2006zs,Sharpe:2014tca} 
predicts that two-dimensional gauge theories with
one-form symmetries will decompose into disjoint unions
of field theories, and we will argue that by
turning on suitable phases in different gerbe sectors, gauging those
one-form symmetries will project onto various components of the
decomposition.  (In passing, see also \cite{Aminov:2019hwg} for another recent
work on one-form symmetries and pure Yang-Mills theories in
two dimensions.)

Consider pure Yang-Mills theories in two dimensions.
Specifically, consider a nonsupersymmetric
pure Yang-Mills theory with gauge group $G$.
Classically, if $K$ denotes the center of $G$,
this theory has a 
$BK = K^{(1)}$ one-form symmetry, which is preserved
in the quantum theory \cite{munsalpriv}.
As a result, since the two-dimensional quantum theory has a 
discrete one-form symmetry, one expects that the theory will
decompose, and indeed, the decomposition of this theory into
$|K|$ gauge theories with discrete theta angles
is discussed in \cite[section 2.4]{Sharpe:2014tca}.

In this section, after briefly reviewing that decomposition,
we will discuss the gauging of that one-form
symmetry, so as to derive various summands in the decomposition,
namely $G/K$ gauge theories with various discrete theta angles.
theory.

First, recall that the partition function of pure bosonic
two-dimensional Yang-Mills theory with gauge group $G$ is known
exactly \cite{Migdal:1975zg,Rusakov:1990rs,Gross:1993hu} and is of the form
\cite[equ'n (3.20)]{Cordes:1994fc}, \cite[equ'n (2.51)]{Witten:1991we}
\begin{equation}
Z \: = \: \sum_R (\dim R)^{2-2g} \exp\left( - A C_2(R) \right),
\end{equation}
where the sum is over representations $R$ of the gauge group $G$,
$g$ is the genus of the two-dimensional spacetime,
and $A$ its area.

If $G$ is not simply-connected, the theory admits discrete theta
angles.  Turning on such discrete theta angles corresponds to
modifying
the sum over representations \cite{Tachikawa:2013hya}.
For example, let us compare the partition functions of $SU(2)$
and $SO(3)_{\pm}$.  For the $SU(2)$ partition function,
one sums over all representations of $SU(2)$, and for the $SO(3)_+$
partition function, one sums over all representations of $SO(3)$ --
meaning, all $SU(2)$ representations that are invariant under the
center.  The $SO(3)_-$ partition function is given by 
\cite{Tachikawa:2013hya} a sum over all $SU(2)$ representations that are not
representations of $SO(3)$.

For the $SU(2)$, $SO(3)_{\pm}$ examples, decomposition is more or less
clear:  the $SU(2)$ partition function is precisely a sum of the
$SO(3)_+$ and $SO(3)_-$ partition functions, 
\begin{equation}
Z(SU(2)) \: = \: Z\left( SO(3)_+ \right) \: + \:
Z\left( SO(3)_- \right).
\end{equation}

This picture generalizes to any pure two-dimensional 
$G$-gauge theory with center $K$,
as discussed in \cite[section 2.4]{Sharpe:2014tca}.
Briefly, 
let $w: {\it Rep} \rightarrow \hat{K}$ denote a map that computes the 
$n$-ality of any given representation.  (It is valued in the characters
$\hat{K}$ of $K$, rather than $K$ itself, as it gives the phase picked up
by a representation of $K \subset G$, hence, is a character of $K$.)
Then, the partition function of the corresponding $G/K$ gauge theory with
discrete theta angle $\lambda \in \hat{K}$ is
\begin{equation}
Z\left( (G/K)_{\lambda} \right) \: = \:
\sum_{R, w(R) = \lambda}  
\left( \dim R \right)^{2-2g} \exp\left( - A C_2(R) \right),
\end{equation} 
where the sum runs over representations $R$ of $G$ of suitable $n$-ality.

In particular, the partition function of the $G$ gauge theory can now
be written 
\begin{eqnarray}
Z(G) & = & \sum_R \left( \dim R \right)^{2-2g} \exp\left( - A C_2(R) \right),
\\
& = & \sum_{\lambda \in \hat{K}} \sum_{R, w(R) = \lambda}  
\left( \dim R \right)^{2-2g} \exp\left( - A C_2(R) \right),
\\
& = & \sum_{\lambda \in \hat{K}} Z\left( (G/K)_{\lambda} \right),
\end{eqnarray}
consistent with decomposition of this two-dimensional theory.

Now, let us consider gauging the center one-form symmetry.
In principle, the partition function has a sum over $K$-gerbes,
weighted by phases, of the form
\begin{equation}
Z( G / BK, \lambda ) \: = \: \frac{1}{|K|} \sum_{z \in K} 
\exp\left( - i \lambda(z) \right) \cdots,
\end{equation}
where the $\cdots$ gives the contribution to the partition function
from sectors that are twisted by the banded $K$ gerbe with characteristic
class $z$.  We will determine those contributions next.

Recall that in an orbifold, in a gerbe-twisted sector,
the usual boundary conditions were twisted:  on $T^2$, for example,
instead of summing over pairs of group elements $g, h$ such that
$gh = hg$, we instead summed over pairs of group elements such that
$gh = hgz$.  Similarly, here, we twist by twisting a cap, when we
construct the partition function by gluing together along a
triangulation.
In other words, ordinarily a cap contributes
(see e.g. \cite[section 3.7]{Cordes:1994fc})
\begin{equation}
Z_{\rm cap}(U) \: = \: \sum_R ( \dim R) \chi_R(U) \exp(-A C_2(R) ),
\end{equation}
where $A$ is the area of the cap, and $U$ is the Wilson line around the
edge.  Here, in a gerbe-$z$-twisted sector, we instead use a twisted
cap, in which the Wilson line on the edge is twisted by $z$:
\begin{equation}
Z_{\rm cap, twisted}(U,z) \: = \:
 \sum_R ( \dim R) \chi_R(z U) \exp(-A C_2(R) ),
\end{equation}
for $z \in K$, the center of the gauge group $G$.
By definition of the function $w$,
\begin{equation}
\chi_R(z U) \: = \: \exp\left( i w(R)(z) \right) \chi_R(U),
\end{equation}
so we can write this as
\begin{equation}
Z_{\rm cap, twisted}(U,z) \: = \:
 \sum_R ( \dim R) \chi_R(U) \exp\left(  i w(R)(z) \right)\exp(-A C_2(R) ).
\end{equation}
We can then compute partition functions in the gerbe-twisted sector
by gluing the twisted cap above into an otherwise normal triangulation.
For example, the partition function of the ordinary (untwisted) theory
on $S^2$ is
\begin{eqnarray}
Z_{S^2} & = & \int d U \, \overline{Z}_{\rm cap}(U) Z_{\rm cap}(U),
\\
& = & \int d U \sum_{R, R'} (\dim R') (\dim R) \overline{ \chi_{R'}(U) }
\chi_R(U) \exp\left( - A({\rm cap}) C_2(R) \right)
\exp\left( - A({\rm cap}') C_2(R') \right),
\nonumber \\ 
& = &
\sum_{R, R'} (\dim R') (\dim R) \delta_{R, R'}
\exp\left( - A({\rm cap}) C_2(R) \right)
\exp\left( - A({\rm cap}') C_2(R') \right),
\\
& = & 
\sum_R (\dim R)^2 \exp\left( - A(S^2) C_2(R) \right),
\end{eqnarray}
the usual partition function in the genus zero case.  In a sector
twisted by the banded $K$ gerbe with characteristic class $z$,
on the other hand, we have instead
\begin{eqnarray}
Z_{S^2}(z) & = & \int d U \, \overline{Z}_{\rm cap}(U) Z_{\rm cap, twisted}(U, z),
\\
& = & \sum_R (\dim R)^2 \exp\left(  i w(R)(z) \right)
\exp\left( - A(S^2) C_2(R) \right),
\end{eqnarray}
following a similar computation, and more generally, on a genus $g$
Riemann surface,
\begin{eqnarray}
Z & = & \sum_R (\dim R)^{2-2g} \exp\left(  i w(R)(z) \right)
\exp\left( - A C_2(R) \right).
\end{eqnarray}

Returning to the partition function of the $BK$-gauged $G$-gauge theory,
and putting the pieces above together,
the resulting partition function is then
\begin{eqnarray}
Z( G / BK, \lambda ) & = &
\frac{1}{|K|} \sum_{z \in K} 
\exp\left( - i \lambda(z) \right)
Z(z), \\
& = &
\frac{1}{|K|} \sum_{z \in K} 
\exp\left( - i \lambda(z) \right)
\sum_R \exp\left( i w(R)(z) \right)
\left( \dim R \right)^{2-2g} \exp\left( - A C_2(R) \right),
\nonumber \\
\end{eqnarray}
where the sum over representations $R$ runs over all representations of $G$.
Now, it is straightforward to check that
\begin{equation}
Z\left( (G/K)_{\lambda} \right) \: = \:
\frac{1}{|K|} \sum_{z \in K} \sum_{R} \exp\left( i (w(R) - \lambda)(z) \right)
\left( \dim R \right)^{2-2g} \exp\left( - A C_2(R) \right),
\end{equation}
as the sum over $z \in K$ effectively performs a projection onto 
representations $R$ such that $w(R) = \lambda$.  As a result, we see that
\begin{equation}
Z( G / BK, \lambda ) \: = \: Z\left( (G/K)_{\lambda} \right),
\end{equation}
and so, at least at the level of partition functions,
gauging the center $BK$ one-form symmetry selects one of the summands
in the decomposition of the original two-dimensional $G$ gauge theory.

\section{Supersymmetric gauge theories}
\label{sect:susygauge}

\subsection{Multiplet for two-form field}

To gauge the one-form symmetry, one needs to introduce a supersymmetrization
of a two-form potential.  A natural candidate superfield is a
twisted chiral superfield, as for example in an abelian gauge theory
in two dimensions, the twisted chiral $\Sigma = \overline{D}_+ D_- V$
depends upon the field strength of the gauge field, but not the gauge field
itself.  For our purposes, following \cite[section 2]{Witten:1993yc}, we take
\begin{eqnarray}
\Sigma^{(2)} & = & \sigma - i \sqrt{2} \theta^+ \overline{\lambda}_+ 
- i \sqrt{2} \overline{\theta}^- \lambda_- +
\sqrt{2} \theta^+ \overline{\theta}^- (D - i B_{01})
- i \overline{\theta}^- \theta^- \partial_- \sigma - i \theta^+ 
\overline{\theta}^+ \partial_+ \sigma
\nonumber \\
& &
 + \sqrt{2} \overline{\theta}^-
\theta^+ \theta^- \partial_- \overline{\lambda}_+
+ \sqrt{2} \theta^+ \overline{\theta}^- \overline{\theta}^+ \partial_+ 
\lambda_- 
- \theta^+ \overline{\theta}^- \theta^- \overline{\theta}^+ 
\partial_+ \partial_- \sigma,
\end{eqnarray}
where $B_{01}$ is the two-form field, and $\sigma, \lambda_{\pm}, D$ are
other fields introduced for the (2,2) supersymmetric completion.
The supersymmetry variations of the components are then given by
\begin{eqnarray}
\delta B_{01} & = & i \overline{\epsilon}_+ \partial_- \lambda_+ 
- i \overline{\epsilon}_- \partial_+ \lambda_-
+ i \epsilon_+ \partial_- \overline{\lambda}_+ -i \epsilon_-
\partial_+ \lambda_-,
\\
\delta \sigma & = & - i \sqrt{2} \overline{\epsilon}_+ \lambda_-
- i \sqrt{2} \epsilon_- \overline{\lambda}_+,
\\
\delta \overline{\sigma} & = & - i \sqrt{2} \epsilon_+ \overline{\lambda}_-
- i \sqrt{2} \overline{\epsilon}_- \lambda_+,
\\
\delta D & = & - \overline{\epsilon}_+ \partial_- \lambda_+ -
\overline{\epsilon}_- \partial_+ \lambda_- +
\epsilon_+ \partial_- \overline{\lambda}_+ +
\epsilon_- \partial_+ \overline{\lambda}_-,
\\
\delta \lambda_+ & = & i \epsilon_+ D + \sqrt{2} \partial_+ \overline{\sigma}
\epsilon_- - B_{01} \epsilon_+,
\\
\delta \lambda_- & = & i \epsilon_- D + \sqrt{2} \partial_- \sigma
\epsilon_+ + B_{01} \epsilon_-,
\\
\delta \overline{\lambda}_+ & = & - i \overline{\epsilon}_+ D
+ \sqrt{2} \partial_+ \sigma \overline{\epsilon}_- - B_{01} \overline{\epsilon}_+,
\\
\delta \overline{\lambda}_- & = & - i \overline{\epsilon}_- D +
\sqrt{2} \partial_- \overline{\sigma} \overline{\epsilon}_+ +
B_{01} \overline{\epsilon}_-.
\end{eqnarray}
Gauge transformations act as 
$\Sigma^{(2)} \mapsto \Sigma^{(2)} + \overline{D}_+ D_- V$, where
\begin{eqnarray}
\overline{D}_+ D_- V & = &
\sigma' - i \sqrt{2} \theta^+ \overline{\lambda}'_+ 
- i \sqrt{2} \overline{\theta}^- \lambda'_- +
\sqrt{2} \theta^+ \overline{\theta}^- (D' - i (dv)_{01})
- i \overline{\theta}^- \theta^- \partial_- \sigma' - i \theta^+ 
\overline{\theta}^+ \partial_+ \sigma'
\nonumber \\
& &
 + \sqrt{2} \overline{\theta}^-
\theta^+ \theta^- \partial_- \overline{\lambda}'_+
+ \sqrt{2} \theta^+ \overline{\theta}^- \overline{\theta}^+ \partial_+ 
\lambda'_- 
- \theta^+ \overline{\theta}^- \theta^- \overline{\theta}^+ 
\partial_+ \partial_- \sigma',
\end{eqnarray}
where
\begin{displaymath}
\delta (dv)_{01} \: = \: 
i \overline{\epsilon}_+ \partial_- \lambda_+ 
- i \overline{\epsilon}_- \partial_+ \lambda_-
+ i \epsilon_+ \partial_- \overline{\lambda}_+ -i \epsilon_-
\partial_+ \lambda_-,
\end{displaymath}
so we see that gauge transformations can be used to eliminate all
components of $\Sigma^{(2)}$ other than $B_{01}$, which undergoes
$B \mapsto B + dv$ for $v$ a one-form.  (This is closely analogous
to Wess-Zumino gauge for ordinary vector superfields, in which gauge
transformations can be used to eliminate some of the fields in the
multiplet.)

\subsection{Abelian examples}  \label{sect:abelexs}

Now, let us apply this to an example.  The prototype for supersymmetric
two-dimensional abelian gauge theories with a
${\mathbb Z}_k^{(1)} =  B {\mathbb Z}_k$ one-form symmetry is the
supersymmetric ${\mathbb P}^n$ model in which all fields have
charge $k > 1$, 
as discussed in \cite{Pantev:2005zs,Pantev:2005rh,Pantev:2005wj}.
(See also \cite{Anber:2018jdf} for a recent related discussion of the
charge $q$ Schwinger model.)

Let us briefly review some basic features of this theory,
beginning with the fact that this description can be consistent with
two different theories.  After all, the perturbative physics of the theory
above and the ordinary supersymmetric ${\mathbb P}^n$ model are identical --
rescaling charges makes no difference.  The difference between these
two theories is nonperturbative in nature, and can be summarized
as follows:
\begin{itemize}
\item If the charges can be rescaled to $1$ to get
the ordinary supersymmetric ${\mathbb P}^n$ model,
then the theta angle has periodicity $2 \pi k$ (in the desciption as
a charge-$k$ model), the axial anomaly breaks $U(1)_A$ to
${\mathbb Z}_{2(n+1)}$, the quantum cohomology ring is 
${\mathbb C}[x]/(x^{n+1}-q)$, and there is no ${\mathbb Z}_k^{(1)}$
one-form symmetry,
\item In the more interesting case, the theta angle has periodicity
$2 \pi$, the axial anomaly breaks $U(1)_A$ to
${\mathbb Z}_{2k(n+1)}$, the quantum cohomology ring is
${\mathbb C}[x]/(x^{k(n+1)} - q)$, and there is a
${\mathbb Z}_k^{(1)}$ one-form symmetry.
\end{itemize}

Now, let us consider gauging that ${\mathbb Z}_k^{(1)}$ one-form symmetry.
The gauge curvature $F_{01}$ appears in the twisted chiral superfield
$\Sigma$, which appears in the action in two terms:
\begin{equation}
\int d^4 \theta \, \Sigma \overline{\Sigma},
\: \: \:
t \int d^2 \tilde{\theta} \, \Sigma,
\end{equation}
i.e. the gauge kinetic term and the FI term.  We can
add the two-form gauge field by replacing $\Sigma$ with
$\Sigma - \Sigma^{(2)}$ in each of the terms above:
\begin{equation}
\int d^4 \theta \, \left( \Sigma - \Sigma^{(2)} \right)
\left( \overline{\Sigma} - \overline{\Sigma}^{(2)} \right),
\: \: \:
t \int d^2 \tilde{\theta} \left( \Sigma - \Sigma^{(2)} \right).
\end{equation}
Now, so as to only gauge the ${\mathbb Z}_k^{(1)}$ symmetry, and not entirely
eliminate the $U(1)$ gauge field, we must
also constraint the two-form field, for which we add the term
\begin{equation}
k \int d^2 \tilde{\theta} \, \Sigma^{(2)} \Lambda,
\end{equation}
where $\Lambda$ is a twisted chiral superfield we introduce with no kinetic
term, to act as a Lagrange multiplier.
This is just a supersymmetrization of a $BF$-type term
\begin{equation}
k \phi B_{01},
\end{equation}
for a scalar $\phi$.

The reader will note that we have not modified the matter kinetic terms,
which contain the gauge field $A_{\mu}$.  Briefly, this will be consistent
so long as the theory admits a ${\mathbb Z}_k^{(1)}$ one-form symmetry.
When we gauge the one-form symmetry, the bundles appearing in the path
integral are `twisted,' which concretely for the gauge field means that
across coordinate patches,
\begin{equation}
A_{\mu} \: \mapsto \: \partial_{\mu} \phi \: + \: \Lambda_{\mu}.
\end{equation}
In other words, in addition to the usual gauge transformation term
$\partial_{\mu} \phi$ for some function $\phi$, the gauge field can also
pick up an affine translation $\Lambda_{\mu}$, determined by the 
${\mathbb Z}_k$ gerbe responsible for the twisting.  
In the present case, because we are
gauging a ${\mathbb Z}_k^{(1)}$ one-form symmetry, the $\Lambda_{\mu}$
are invisible to the charge-$k$ matter fields, and so the kinetic terms
are unmodified.  (Put another way, after gauging the ${\mathbb Z}_k^{(1)}$,
the original gauge field $A$ is no longer a well-defined ordinary
gauge field, but $k A$ is, and because the matter fields all have charge
$k$, only the product $kA$ appears in the kinetic terms.)

If the matter was not invariant under that
${\mathbb Z}_k^{(1)}$ one-form symmetry, then we would need to add
explicit couplings to the matter kinetic energies to make them 
well-defined across coordinate patches, because of those
affine translations $\Lambda_{\mu}$.  Put another way,
only in a theory with a ${\mathbb Z}_k^{(1)}$ one-form symmetry
does it suffice to only modify the terms in the supersymmetric
action involving $\Sigma$.

Note in passing that the worldsheet theta angle flows in the IR
to the pullback of of the target-space $B$ field, which is distinct from
the worldsheet $B$ field we introduce to gauge the one-form symmetry.
As a result, we cannot gauge the one-form symmetry by merely
promoting the worldsheet theta angle to an axion.

Finally, to allow for the gerbes in the path integral to contribute with
different phases, we also add a term
\begin{equation}
\alpha \int d^2 \tilde{\theta} \, \Sigma^{(2)}.
\end{equation}
Since $\Sigma^{(2)}$ is constrained to define a ${\mathbb Z}_k$
BF theory, $\alpha$ is only meaningful mod $k$.

In particular, in the analogue of Wess-Zumino gauge, 
\begin{equation}
\int d^2 \tilde{\theta} \, \Sigma^{(2)} \: \propto \: B_{01},
\end{equation}
and so adding this term to the Lagrangian introduces a gerbe-dependent phase
into the path integral, weighting different worldsheet gerbes differently.

Briefly, in the case of the worldsheet ${\mathbb P}^n$ model with
charges $k > 1$, gauging the one-form symmetry as above has the effect
of changing the size of the $U(1)$ by a factor of $k$.  One way to understand
this is in terms of its effect on the theta angle periodicity.
The theory with the ${\mathbb Z}_k^{(1)}$ one-form symmetry
has a theta angle that is $2 \pi$ periodic.  When we gauge the
one-form symmetry, we sum over ${\mathbb Z}_k$ gerbes, which twist
the bundle to a `fractional' bundle.  Such fractional bundles have
fractional Chern classes; for example, over
${\mathbb P}^1_{[k,k]}$, one can have line bundles
\begin{equation}
{\cal O}(m) \: \longrightarrow \: {\mathbb P}^1_{[k,k]},
\end{equation}
for any integer $m$, including integers less than $k$, and such a line
bundle has $c_1 = m/k$.
(See e.g. \cite{Anderson:2013sia} for a discussion of bundles on gerbes.)
As a result, in the new theory, the theta angle multiplies a factor
$\int F$ which is valued in multiples of $1/k$ instead of integers,
hence in this theory the theta angle is now $2 \pi k$ periodic.
Such $2 \pi k$ periodicity signals that the resulting theory is
equivalent to the ordinary ${\mathbb P}^n$ model.

Alternatively, one can think of gauging the one-form symmetry as
a change in the `size' of the $U(1)$, replacing $U(1)$ by
$U(1)/{\mathbb Z}_k$, for which charge $k$ fields become
charge $1$ fields, restoring the original theory.

The axial $U(1)_A$ anomaly is of the form
\begin{equation}
[ D \psi ] \: \mapsto \: [D \psi] \exp\left( i 2 k (n+1) \omega
Q_{\rm top} \right),
\end{equation}
where 
\begin{equation}
Q_{\rm top} \: = \: \frac{1}{2 \pi} \int d^2 x F_{01},
\end{equation}
and for the original ${\mathbb P}^n$ model,
$Q_{\rm top} \in {\mathbb Z}$.
As a result, there is an anomaly-free ${\mathbb Z}_{2k(n+1)}
\subset U(1)_A$ subgroup corresponding to phases in which
$\omega \in 2 \pi {\mathbb Z} /  (2 k (n+1) )$.
If we gauge the ${\mathbb Z}_k^{(1)}$ one-form symmetry,
then $Q_{\rm top} \in {\mathbb Z}/k$.
Now, if we restrict to the anomaly-free subgroup of $U(1)_A$ of
the original theory, before gauging ${\mathbb Z}_k^{(1)}$,
then 
\begin{equation}
\omega \: = \: \frac{2 \pi m}{2k(n+1)}
\end{equation}
for $m \in {\mathbb Z}$,
and in the new theory, after gauging the ${\mathbb Z}_k^{(1)}$,
\begin{equation}
[D \psi] \: \mapsto \: [D \psi] \exp\left( 2 \pi i \frac{m}{k} \right).
\end{equation}
Thus, since the original anomaly-free ${\mathbb Z}_{2k(n+1)} \subset
U(1)_A$ now becomes anomalous after gauging the one-form symmetry,
we say that there is a mixed 0-form / 1-form anomaly, mixing
the ${\mathbb Z}_{2k(n+1)}$ axial 0-form symmetry and the
${\mathbb Z}_k^{(1)}$ one-form symmetry of the original theory.
Specifically, the anomaly is given by the phase
\begin{equation}
\exp\left( 2 \pi i / k \right).
\end{equation}

Another interpretation of this mixed 0/1-form anomaly is as follows.
Note that in the new theory, under an axial $U(1)_A$ rotation,
\begin{equation}
[D \psi] \: \mapsto \: [D \psi] \exp\left(i 2 (n+1) \omega m \right),
\end{equation}
where $m \in {\mathbb Z}$, $m = k Q_{\rm top}$.
As a result, after gauging the one-form symmetry of the original theory,
the axial anomaly of the new theory has a nonanomalous
${\mathbb Z}_{2(n+1)}$ subgroup, matching the standard result for
the axial anomaly in the ordinary ${\mathbb P}^n$ model.

This is only one generalization of the supersymmetric 
${\mathbb P}^n$ model.  A more general family can be described
as follows \cite{Pantev:2005zs}.  (See also e.g. \cite{Pantev:2005zs} 
for analogous generalizations of
GLSMs for toric varieties.)
Consider a GLSM with gauge group $U(1)^2$ and matter fields with
charges as follows:
\begin{center}
\begin{tabular}{ccccc}
 & $x_0$ & $\cdots$ & $x_n$ & $z$ \\ \hline
$U(1)_{\lambda}$ & $1$ & $\cdots$ & $1$ & $-m$ \\
$U(1)_{\mu}$ & $0$ & $\cdots$ & $0$ & $k$
\end{tabular}
\end{center}
Here, we interpret $x_{0, \cdots, n}$ in terms of homogeneous
coordinates on ${\mathbb P}^n$, as before.  We interpret
$z$ as an analogous coordinate on the total space of
${\cal O}(-m)$; however, from the D-term associated to $U(1)_{\lambda}$
for nonzero FI parameter,
we omit the zero locus, and then gauge out ${\mathbb C}^{\times}$
rotations.  

Geometrically, this GLSM describes a family of ${\mathbb Z}_k$
gerbes over ${\mathbb P}^n$, i.e., generalized spaces with
${\mathbb Z}_k^{(1)}$ one-form symmetries.  Such gerbes are
classified by a characteristic class in $H^2({\mathbb P}^n,
{\mathbb Z}_k)$, and GLSMs of the form above describe all possible
${\mathbb Z}_k$ gerbes over ${\mathbb P}^n$.  The GLSM with the
charges shown describes the gerbe of characteristic class $-m$ mod $k$,
for example, and the previous generalization of the 
${\mathbb P}^n$ model corresponds to the gerbe of
characteristic class $-1$ mod $k$.
The characteristic class is reflected in the quantum cohomology
ring, for example, which is given by
\begin{equation}
{\mathbb C}[x,y] / \langle y^k - 1, x^{n+1} y^{-m} - q \rangle.
\end{equation}

Gauging the one-form symmetry acts here much as in the previous case:
the charge $k$ field becomes, in effect, a charge 1 field,
the global ${\mathbb Z}_k^{(1)}$ disappears, and the theory reduces to
a copy of the ordinary ${\mathbb P}^n$ model.

Other examples of gerbe structures arising in GLSMs are
described in \cite{Pantev:2005zs}.  Briefly, gauging one-form symmetries
there follows the same pattern as outlined here for these
various analogues of the supersymmetric
${\mathbb P}^n$ model.

In passing, it is also worth mentioning that gauging the flavor symmetry
in these models can also be of interest, see
\cite[section 2.4]{Komargodski:2017dmc} and
\cite{Tanizaki:2017qhf,Tanizaki:2018xto,Misumi:2019dwq}.

\subsection{Mirrors to two-dimensional gauge theories}

So far we have discussed one-form symmetries in two-dimensional gauge
theories, next we shall discuss how the one-form symmetries are
visible in the mirrors, following \cite{Pantev:2005zs,Hellerman:2006zs}.

\subsubsection{Abelian mirrors}

For our first generalization of the supersymmetric
${\mathbb P}^n$ model, discussed in section~\ref{sect:abelexs},
a $U(1)$ gauge theory with charges $k > 1$, the mirror was computed
in \cite{Pantev:2005zs} and is a Landau-Ginzburg model with
a superpotential of the form
\begin{equation}
W \: = \: \exp(-Y_1) + \cdots + \exp(-Y_n) +
q \Upsilon \prod_{i=1}^n \exp(+Y_i),
\end{equation}
where $\Upsilon$ is a ${\mathbb Z}_k$-valued field.  (This is clearly
equivalent to a disjoint union of Landau-Ginzburg models,
and is one way to understand decomposition in this example.)
For more general gerbes over ${\mathbb P}^n$, mirrors were
also computed in \cite{Pantev:2005zs}, and have the form
\begin{equation}
W \: = \: \exp(-Y_1) + \cdots + \exp(-Y_n) +
q \Upsilon^{-m} \prod_{i=1}^n \exp(+Y_i).
\end{equation}

We can see the axial anomaly as follows.
First, for simplicity, consider the mirror to the ordinary
supersymmetric ${\mathbb P}^n$ model, given by a 
Landau-Ginzburg model with superpotential
\begin{equation}
W \: = \: \exp(-Y_1) + \cdots + \exp(-Y_n) +
q  \prod_{i=1}^n \exp(+Y_i).
\end{equation}
The mirror to the axial symmetry of the original theory is the
translation 
\begin{equation}  \label{eq:axial-mirror}
Y_i \: \mapsto \: Y_i - 2 i \alpha,
\end{equation}
where $\alpha$ parametrizes the $U(1)_R$.
Under this translation, the superpotential terms $\exp(-Y_i)$ have
weight 2, as needed for an R-symmetry.  The anomaly is visible in the
last term.  The product
\begin{equation}
\prod_{i=1}^n \exp(+Y_i)
\end{equation}
has weight $-2n$, so in order for the corresponding term to 
have the same weight as the other terms, we must require that
$q$ have weight $2+2n = 2(n+1)$.  This reflects the fact that
$q$ encodes the theta angle of the original theory, which will effectively
shift under an anomalous chiral rotation.  We can read off the anomaly
from the theta angle shift, and see in particular there there is a 
nonanomalous ${\mathbb Z}_{2(n+1)}$ subgroup.

It is straightforward to modify this computation to apply to the
gerbe mirror case.  There, again the mirror of the axial symmetry is
the translation~(\ref{eq:axial-mirror}), and so the terms $\exp(-Y_i)$
have weight two, as needed for an R-symmetry.  
To understand the action on the remaining terms, first recall that for
the first gerbe, ${\mathbb P}^n_{[k,k,\cdots,k]}$, the theta angle
of the original theory is encoded in $\tilde{q}$, which satisfies
\begin{equation}
\tilde{q} \: = \: \exp(-k Y_{n+1}) \prod_{i=1}^n \exp(-k Y_i ),
\end{equation}
and so transforms as
\begin{equation}   \label{eq:abelmirror:tildeq}
\tilde{q} \: \mapsto \: \tilde{q} \exp\left( + 2 k (n+1) i \alpha \right).
\end{equation}
The combination $q \Upsilon = \tilde{q}^{1/k}$, where the $\Upsilon$ reflects
the ambiguity by $k$th roots of unity, and so we see that under the
mirror of the axial symmetry,
\begin{equation}
q \Upsilon \: \mapsto \: q \Upsilon \exp\left( + 2 (n+1) i \alpha \right).
\end{equation}
This describes the mirror of the axial anomaly.
The more general gerbe mirror can be analyzed similarly.

Let us now turn our attention to gauging the ${\mathbb Z}_k^{(1)}$ one-form
symmetry in the mirror.  In fact, this is extremely easy:  since the
one-form symmetry is realized just by changing $\Upsilon$ from
one $k$th root of unity to another, gauging the one-form symmetry
gauges `translations' of this discrete-valued field, effectively
removing it from the theory.  In this language, it is also easy to
see the mixed 0/1-form anomaly in the mirror.  In the original mirror,
from equation~(\ref{eq:abelmirror:tildeq}), there is an anomaly-free
${\mathbb Z}_{2k(n+1)}$ subgroup of the axial $U(1)_A$, defined by
\begin{equation}
\alpha \: \in \: \frac{2 \pi {\mathbb Z}}{ 2 k (n+1) },
\end{equation}
which leaves $\tilde{q}$ invariant.
However, after gauging the ${\mathbb Z}_k^{(1)}$ (meaning, removing
$\Upsilon$ from the mirror), such $\alpha$ shifts $q$ by a factor of
\begin{equation}
\exp\left( 2 (n+1) \frac{2 \pi i {\mathbb Z} }{2 k (n+1)} \right)
\: = \: 
\exp\left( \frac{2 \pi i {\mathbb Z} }{k} \right),
\end{equation}
and so we see that there is a mixed 0/1-form anomaly, given by the phase
$\exp\left( 2 \pi i / k \right)$, matching the result for the original
theory.
The more general gerbe mirror can be analyzed similarly.

\subsubsection{Nonabelian mirrors}

An extension of the Hori-Vafa ansatz for mirrors to abelian GLSMs was
proposed in \cite{Gu:2018fpm,Chen:2018wep,Gu:2019zkw,Gu:2019byn}.
For example, \cite[section 12]{Gu:2018fpm} discussed pure (2,2) supersymmetric
$SU(k)$ gauge theories in two dimensions, which have a ${\mathbb Z}_k^{(1)}$
one-form center symmetry.  

Let us first review the structure of those
mirrors.
The mirror to the pure supersymmetric $SU(2)$ theory was described
there by a Landau-Ginzburg orbifold with superpotential
\begin{equation}
W \: = \: 2 \Sigma \ln \left( \frac{ X_{12} }{X_{21}} \right)
\: + \: X_{12} \: + \: X_{21},
\end{equation}
the mirror to the pure $SO(3)_+$ theory was a closely-related
Landau-Ginzburg orbifold with superpotential
\begin{equation}
W \: = \: \Sigma \ln \left( \frac{ X_{12} }{X_{21}} \right)
\: + \: X_{12} \: + \: X_{21},
\end{equation}
and the mirror to the pure $SO(3)_-$ theory was the Landau-Ginzburg
orbifold with superpotential
\begin{equation}
W \: = \: \Sigma \ln \left( \frac{ X_{12} }{X_{21}} \right)
\: + \: \pi i \Sigma \: + \: X_{12} \: + \: X_{21}.
\end{equation}
When solving for the vacua in these mirrors, one finds a continuum of
vacua in both the
$SU(2)$ and $SO(3)_-$ theories, corresponding to $X_{12} = - X_{21}$,
but no vacua at all in the $SO(3)_+$ theory, consistent with the
expected decomposition \cite{Sharpe:2014tca}
\begin{equation}
SU(2) \: = \: SO(3)_+ \: + \: SO(3)_-,
\end{equation}
and analogous decompositions in other pure supersymmetric gauge theories.

Ultimately, the presence of the ${\mathbb Z}_2^{(1)}$ one-form symmetry
in the mirror to the $SU(2)$ theory is reflected in the factor of $2$
in the first term in the superpotential, just as the
mirror to the ${\mathbb Z}_k^{(1)}$ one-form symmetry in the abelian
examples of the last section was reflected in a superpotential
term of the form $k \Sigma Y$, which forced $\exp(-Y)$ to be a $k$th root of
unity.  Here, because of the factor of $2$,
one requires
\begin{equation}
\left( \frac{ X_{12} }{ X_{21} } \right)^2 \: = \: 1,
\end{equation}
instead of merely
\begin{equation}
\left( \frac{ X_{12} }{ X_{21} } \right) \: = \: \pm 1,
\end{equation}
and so in the $SU(2)$ mirror one sums over the two roots of the equation
above, which are easily seem to correspond to the two $SO(3)$ mirrors for
either discrete theta angle.

Briefly, much as in the abelian case, gauging the
${\mathbb Z}_2^{(1)}$ of the original theory corresponds to removing that
choice of roots -- removing the factor of $2$ in the $SU(2)$ mirror
superpotential -- and possibly adding a phase, to distinguish the 
$SO(3)_+$ from $SO(3)_-$ mirrors.

\subsection{Partition functions of abelian and nonabelian theories}

Partition functions of two-dimensional
supersymmetric gauge theories on $S^2$ were
computed in \cite{Benini:2012ui,Doroud:2012xw}, and have the form
\cite[equ'n (3.34)]{Benini:2012ui}
\begin{equation}   \label{eq:part-fn}
Z_{S^2} \: = \: \frac{1}{|{\cal W}|} \sum_{\mathfrak m} 
\int \left( \prod_j \frac{d \sigma_j}{2\pi} \right) Z_{\rm class}(
\sigma, {\mathfrak m}) Z_{\rm gauge}(\sigma, {\mathfrak m})
\prod_{\Phi} Z_{\Phi}(\sigma, {\mathfrak m}; \tau, {\mathfrak n}) ,
\end{equation}
where \cite[equ'n (3.35)]{Benini:2012ui}
\begin{eqnarray*}
Z_{\rm class}(\sigma,{\mathfrak m}) 
& = & e^{-4 \pi i \xi {\rm Tr}\, \sigma - i \theta
{\rm Tr} \, {\mathfrak m}} \exp\left( 8 \pi i r
{\rm Re}\, \tilde{W}(\sigma/r + i {\mathfrak m}/(2r)) \right), \\
Z_{\rm gauge}(\sigma,{\mathfrak m}) & = & 
\prod_{\alpha \in G} \left( \frac{ |\alpha( 
{\mathfrak m})| }{2} \: + \: i \alpha(\sigma) \right)
\: = \:
\prod_{\alpha > 0} \left( \frac{ \alpha({\mathfrak m})^2}{4} \: + \:
\alpha(\sigma)^2 \right), \\
Z_{\Phi}(\sigma, {\mathfrak m}; \tau, {\mathfrak n}) & = & 
\prod_{\rho \in R_{\Phi}} \frac{
\Gamma\left( \frac{R[\Phi]}{2} \: - \: i \rho(\sigma) \: - \:
i f^a[\Phi] \tau_a \: - \: \frac{
\rho({\mathfrak m}) + f^a[\Phi] n_a }{2} \right)
}{
\Gamma\left( 1 \: - \: \frac{R[\Phi]}{2} \: + \: i \rho(\sigma)
\: + \: i f^a[\Phi] \tau_a \: - \: \frac{
\rho({\mathfrak m}) + f^a[\Phi]n_a }{2} \right)
} .
\end{eqnarray*}
Briefly, in the notation of \cite{Benini:2012ui},
$R[\Phi]$ is the R-charge of a chiral multiplet $\Phi$,
$f^a[\Phi]$ a non-R-charge,
$R_{\Phi}$ denotes the gauge group representation in which
$\Phi$ appears, ${\cal W}$ is the Weyl group of the gauge group,
and
$\tau = (\tau_a)$ and ${\mathfrak n}=(n_a)$ define twisted masses for the
chiral superfield.
The ${\mathfrak m}$ are elements of the 
cocharacter 
(dual weight) lattice for the
gauge group, meaning for any weight $\rho$ in the weight lattice,
$\rho({\mathfrak m}) \in {\mathbb Z}$.  

Decomposition of two-dimensional supersymmetric gauge theories with
center-invariant matter at the level of such partition functions was
discussed in \cite{Sharpe:2014tca}.
Let us briefly review that analysis here.
(We will implicitly assume in this section that the center one-form
symmetry is unbroken by any sort of anomalies.)

For simplicity, we begin with a comparison of two-dimensional supersymmetric
$SU(2)$
gauge theories with center-invariant matter to corresponding $SO(3)_{\pm}$
gauge theories.  As discussed in \cite{Sharpe:2014tca}, the cocharacter
lattice of $SU(2)$ is twice as large as that of $SO(3)$, so if
for $SO(3)$, ${\mathfrak m}$ varies over all integers,
then for $SU(2)$, ${\mathfrak m}$ varies over even integers.
Furthermore, the discrete theta angle of the $SO(3)$ theories is
encoded in the partition function lattice sum as a factor
\begin{equation}
\exp\left( - i \pi {\mathfrak m} \right) \: = \: 
(-)^{\mathfrak m}.
\end{equation}
At the level of partition functions, decomposition can now be
understood as follows.  If we write the partition function of 
a $SO(3)_+$ gauge theory with center-invariant matter in the form
\begin{equation}
Z\left( SO(3)_+ \right) \: = \: \frac{1}{2}
\sum_{ {\mathfrak m} \in {\mathbb Z}} A({\mathfrak m}),
\end{equation}
for some function $A({\mathfrak m})$ given by the supersymmetric
localization formulas earlier, then for the
corresponding $SO(3)_-$ gauge theory,
\begin{equation}
Z\left( SO(3)_- \right) \: = \: \frac{1}{2}
\sum_{ {\mathfrak m} \in {\mathbb Z}} (-)^{\mathfrak m}A({\mathfrak m}),
\end{equation}
and for the corresponding $SU(2)$ gauge theory,
\begin{equation}
Z\left( SU(2) \right) \: = \:
\sum_{ {\mathfrak m} \in 2 {\mathbb Z} } A({\mathfrak m})
\: = \:
Z\left( SO(3)_+ \right) \: + \:
Z\left( SO(3)_- \right).
\end{equation}
In particular, the form of the expressions from supersymmetric
localization is the same for the three theories
$SO(3)_{\pm}$ and $SU(2)$, hence the function $A({\mathfrak m})$
is necessarily the same in each case; only the lattice sum,
and a possible phase factor, can differ.

In this spirit, we can now gauge the ${\mathbb Z}_2^{(1)}$ one-form
symmetry at the level of partition functions as follows.
As before, we sum over banded ${\mathbb Z}_2$ gerbes on the worldsheet.
For the trivial ${\mathbb Z}_2$ gerbe, we then sum over the lattice
of the original $SU(2)$ gauge theory (meaning,
${\mathfrak m} \in 2 {\mathbb Z}$).  For the nontrivial 
${\mathbb Z}_2$ gerbe, we sum over $SU(2)$ bundles twisted by the
${\mathbb Z}_2$ gerbe, which is to say, $SO(3)$ bundles which are
not also $SU(2)$ bundles,
and hence we sum over odd ${\mathfrak m}$.
We add a gerbe-dependent phase factor $\epsilon$.  Thus, 
using ``$SU(2)/B{\mathbb Z}_2$'' to denote the $SU(2)$ gauge theory
with the action of $B {\mathbb Z}_2 = {\mathbb Z}_2^{(1)}$ gauged,
we have
\begin{equation}
Z\left( SU(2) / B {\mathbb Z}_2 \right) 
\: = \:
\frac{1}{2} \epsilon(1) \sum_{ {\mathfrak m} \in 2 {\mathbb Z} } 
A({\mathfrak m}) \: + \:
\frac{1}{2} \epsilon(-1) \sum_{ {\mathfrak m} \in 2{\mathbb Z}+1 }
A({\mathfrak m}),
\end{equation}
where we take $\epsilon(1) = 1$.
In the case that $\epsilon(-1) = 1$, we have
\begin{equation}
Z\left( SU(2) / B {\mathbb Z}_2 \right) 
\: = \:
\frac{1}{2} \sum_{ {\mathfrak m} \in  {\mathbb Z} } A({\mathfrak m})
\: = \: Z\left( SO(3)_+ \right).
\end{equation}
In the case that $\epsilon(-1)=-1$, we have
\begin{equation}
Z\left( SU(2) / B {\mathbb Z}_2 \right) 
\: = \:
\frac{1}{2} \sum_{ {\mathfrak m} \in  {\mathbb Z} }
(-)^{\mathfrak m}  A({\mathfrak m})
\: = \: Z\left( SO(3)_- \right).
\end{equation}

Thus, we see that the two different gaugings of 
$B{\mathbb Z}_2 = {\mathbb Z}_2^{(1)}$ yield the two different discrete
theta angles, the two theories $SO(3)_{\pm}$.

Now, we shall generalize this to more general semisimple
gauge groups $G$, following \cite[section 2.6]{Sharpe:2014tca}.
Let $K$ denote the center of $G$ and $M_G$ the cocharacter lattice of
$G$, so that $M_G \subset M_{G/K}$ and $M_{G/K} / M_G$ has as many
element sas $K$.  The integral of the analogue of the second
Stiefel-Whitney class is encoded in the map $w$ in
\begin{equation}
1 \: \longrightarrow \: M_G \: \longrightarrow \: M_{G/K} \:
\stackrel{w}{\longrightarrow} \: K \: \longrightarrow \: 1.
\end{equation}
Then,
\begin{equation}
\frac{1}{|K|} \sum_{\mu \in \hat{K}} \exp\left( i \mu( w({\mathfrak m}) ) 
\right)
\end{equation}
is a projection operator that projects $M_{G/K}$ onto $M_G$.
In this language, the partition function of a two-dimensional
supersymmetric $G$-gauge theory can be expressed in the form
\begin{eqnarray}
Z(G) & = & \sum_{ {\mathfrak m} \in M_G } A( {\mathfrak m} ),
\\
& = & \frac{1}{|K|} \sum_{\lambda \in \hat{K} } 
\sum_{ {\mathfrak m} \in M_{G/K} } \exp\left( i \lambda(
w({\mathfrak m}) ) \right) A({\mathfrak m}),
\\
& = & \sum_{\lambda \in \hat{K} } Z\left( (G/K)_{\lambda} \right),
\end{eqnarray}
where
\begin{equation}
Z\left( (G/K)_{\lambda} \right)
\: = \: \frac{1}{|K|} \sum_{ {\mathfrak m} \in M_{G/K} }
\exp\left( i \lambda(
w({\mathfrak m}) ) \right) A({\mathfrak m})
\end{equation}
is the partition function of the corresponding
$G/K$ gauge theory with discrete theta angle $\lambda$.
In this fashion, we see how, at the level of partition functions,
a two-dimensional supersymmetric $G$ gauge theory with center-invariant
matter decomposes into a disjoint union of $G/K$ gauge theories
with discrete theta angles.

Now, let us gauge the $BK = K^{(1)}$ one-form symmetry of this
theory.
Proceeding as before, to gauge the action of $BK$ on the
$G$-gauge theory, the partition function is a sum over
$K$ gerbes
\begin{equation}
\frac{1}{|K|} \sum_{z \in K} \cdots
\end{equation}
in which for each $K$ gerbe, one sums over gerbe-twisted $G$
bundles, which in this case means $G/K$ bundles defined by
cocharacters ${\mathfrak m} \in M_{G/K}$ which are not also
$G$ bundles, meaning ${\mathfrak m} \not\in M_G \subset M_{G/K}$.
Furthermore, to be twisted by $z$ specifically, as opposed to a random
element of $K$, we also need to require that the bundles twisted by
the $K$-gerbe with characteristic class $z$ have $w({\mathfrak m}) = z$.

Putting this together, we see that the partition function
of the $G$ gauge theory after gauging by $BK$ has the form
\begin{eqnarray}
Z( G / BK ) & = & \frac{1}{|K|} \sum_{z \in K}
\epsilon(z)
\sum_{ {\mathfrak m} \in M_{G/K}, w({\mathfrak m}) = z }
A( {\mathfrak m} ),
\end{eqnarray}
where $\epsilon(z)$ represent phases introduced for each gerbe sector.
We can relate $\epsilon(z)$ to a particular character $\lambda \in
\hat{K}$ by taking
\begin{equation}
\epsilon(z) \: = \: \exp\left( i \lambda(z) \right).
\end{equation}
Then, corresponding to that character $\lambda$, we have the
partition function
\begin{eqnarray}
Z( G / BK, \lambda) & = & 
\frac{1}{|K|} \sum_{z \in K}
\sum_{ {\mathfrak m} \in M_{G/K}, w({\mathfrak m}) = z }
\exp\left( i \lambda( w( {\mathfrak m} ) ) \right)
A( {\mathfrak m} ),
\\
& = &
\frac{1}{|K|} \sum_{ {\mathfrak m} \in M_{G/K} }
\exp\left( i \lambda( w( {\mathfrak m} ) ) \right)
A( {\mathfrak m} ),
\\
& = & Z\left( (G/K)_{\lambda} \right).
\end{eqnarray}

Thus, we see that the partition function of the $G$-gauge theory,
with gauged $BK$ action determined by $\lambda \in \hat{K}$,
matches the partition function of the corresponding
$G/K$ gauge theory with discrete theta angle determined by
$\lambda$, as expected.

\section{K theory}
\label{sect:Ktheory}

One of the consistency checks of decomposition applied in
\cite{Hellerman:2006zs} involved computing D-brane charges in
gauge theories with trivially-acting subgroups -- namely,
K theory classes.  Although a subgroup of the gauge (or orbifold) group
may act trivially on the underlying space, it might still act
nontrivially on bundles over the space, which leads to a decomposition
of K theory classes parametrized by irreducible representations of the
trivially-acting subgroup, or put another way, a decomposition of K theory
into K theory groups of the various summands appearing in decomposition.
Thus, briefly, D-brane charges decompose in the fashion predicted
by decomposition; schematically,
\begin{equation}
K \: = \: K\left( \coprod_i X_i \right).
\end{equation}

We can similarly understand how gauging a one-form symmetry can select
out summands in K theory.  To be clear, consider a $G$-gauged nonlinear
sigma model on $X$, where $K \subset G$ acts trivially, a $K$ gerbe
on $[X/H]$ for $H = G/K$.  For simplicity, assume that the $K$ gerbe
is banded, and $K$ abelian.  When we gauge the $BK = K^{(1)}$ one-form
symmetry, we will see momentarily that
it acts in the worldsheet theory
by multiplying the various K theory elements by
phases.  

\begin{figure}[h]
\centering
\begin{picture}(100,100)(0,0)
\CArc(50,50)(50,0,360)
\DashLine(50,50)(100,50){5}
\Text(75,54)[b]{$z$}
\end{picture}
\caption{Open string disk diagram with bulk twist field insertion.}
\label{fig:disk}
\end{figure}
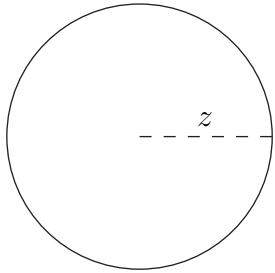

Physically, we can see the details of this operation as follows.
Consider an open string disk diagram.  The 2-group $BK$ on the Chan-Paton
factors by `tensoring' the corresponding bundle by a principal $K$ bundle.
If the disk diagram has a $K$-twist field in its bulk,
as illustrated in figure~\ref{fig:disk}.

Because the disk is contractible, principal $K$ bundles on the disk are
completely determined by their boundary conditions, here by their
holonomy around the edge of the disk.  A bulk twist field in the interior
creates a `branch cut' by $z \in K$, hence the worldsheet bundle
on the disk diagram shown in figure~\ref{fig:disk}.  
When acting on the Chan-Paton factors on the boundary of this disk,
the action of $BK$ on a K theory class in a component of the decomposition
indexed by character $\rho \in \hat{K}$, is to multiply it by a phase
$\rho(z)$.

Putting these pieces together, if in the bulk of the disk one
inserts a projection operator of the form
\begin{equation}
\sum_{z \in K} \epsilon(z) {\cal O}_z,
\end{equation}
where ${\cal O}_z$ is a twist field corresponding to $z$,
the effect is to make the corresponding disk amplitude
vanish if the Chan-Paton factors are not associated with a particular
character of $K$.  For example, if the phases $\epsilon(z)$ are all
$1$, then the effect is to project onto K theory classes that are
associated with the trivial representation of $K$.
The result is consistent with the picture presented elsewhere in this paper,
that gauging one-form symmetries in two-dimensional theories projects
onto components of decomposition.

\section{Conclusions}

In this paper we have discussed the gauging of one-form symmetries in
two-dimensional theories, and how this selects out summands in
the decomposition of such two-dimensional theories.  We have in particular
tracked through the examples of two-dimensional theories with one-form
symmetries discussed in 
\cite{Pantev:2005zs,Pantev:2005rh,Pantev:2005wj,Sharpe:2014tca},
a mix of orbifolds and two-dimensional nonsupersymmetric and supersymmetric
gauge theories, and explicitly demonstrated that gauging the one-form
symmetries reverses decomposition, by selecting out one component in the 
summand.  In so doing, we have given a very concrete description of the
topological configurations over which path integrals sum when gauging
one-form symmetries, and also discussed the available patterns of
one-form symmetries in disjoint unions.

We have primarily focused on two-dimensional theories corresponding to
banded gerbes.  We have outlined some results for nonbanded and nonabelian
gerbes, but leave a thorough examination of those cases for future work.

\section{Acknowledgements}

We would like to thank D.~Benzvi and T.~Pantev for many useful discussions.
We would also like to thank A.~Cherman, M.~\"{U}nsal, and Y.~Tanizaki for many
useful discussions and hospitality during a visit to NCSU in fall 2019.
In addition, we would like to thank E.~Poppitz for many useful discussions
during summer 2019 at the Aspen Center for Physics, which is supported
by NSF grant PHY-1607611.
E.S. was partially supported by NSF grant PHY-1720321.

\appendix

\section{Stacks and 2-stacks}
\label{app:stacks}

An elegant way to understand gauge theories and quotients by ordinary
groups is in terms of a stack, a generalization of a space,
for which the prototypical example is
$[X/G]$, for $G$ acting on a space $X$.  Mathematical discussions of
these can be found in e.g. \cite{gomez,lmb,bx,hein,metz,noohi1,noohi2,noohi3}.  
Stacks admit metrics, spinors, and so
forth, and so one can reasonably expect that one can define a nonlinear
sigma model on a stack, 
see e.g. \cite{Pantev:2005zs,Pantev:2005rh,Pantev:2005wj} for details.

To similarly make sense of gauging one-form symmetries,
in principle one should appeal to 2-stacks.

Now, an ordinary stack can be defined by its incoming maps from
other spaces (a description that is very relevant for sigma models).  
For example, for $G$ finite, maps $\Sigma \rightarrow [X/G]$ 
are defined by pairs consisting of
\begin{itemize}
\item a principal $G$ bundle $E \rightarrow \Sigma$,
\item a $G$-equivariant map ${\rm Tot}(E) \rightarrow X$,
\end{itemize}
which can straightforwardly be seen to correspond to the twisted
sectors one sums over in a path integral description of orbifolds.

In principle, a 2-stack $[{\mathfrak X}/BG]$, where ${\mathfrak X}$ is
an ordinary stack admitting an action of $BG$.  
A map $\Sigma \rightarrow [{\mathfrak X}/BG]$ should be again defined by
data including a $G$-gerbe on $\Sigma$.

In this language, we can get a geometric picture of what gauging a
one-form symmetry is accomplishing.  Suppose we start with a $G$-gerbe
on a space $X$, for $G$ abelian.  The $G$-gerbe is the total space of
a $BG$ bundle on $X$.  (As such, a gerbe is a `generalized
space' with a one-form symmetry, which is the basic reason why
sigma models on gerbes have global one-form symmetries.)  
Gauging $BG$ is just quotienting those $BG$
fibers, leaving (modulo details of two-group actions) the underlying space
$X$.  For example, for a trivial $G$-gerbe on $X$, the gerbe is
$[X/G] = X \times BG$, and so gauging the one-form symmetry $G^{(1)} = BG$
is just the quotient
\begin{equation}
\left[ \frac{ X \times BG}{BG} \right] \: \cong \: X.
\end{equation}
This simple formal observation gives a mathematical prototype for
the constructions described in this paper.

\end{document}